\documentstyle[11pt]{article}
\begin{document}
\begin{large}
\begin{titlepage}

\vspace{0.2cm}

\title{Pair production of the lightest chargino via gluon-gluon collisions
\footnote{The project supported by National Natural Science
          Foundation of China}}
\author{{ Ma Wen-Gan$^{a,b,c}$, Du Fei$^{b}$, Zhou Mian-Lai$^{b}$,
          Jiang Yi$^{b}$, Han Liang$^{b}$ and Zhou Hong$^{b}$}\\
{\small $^{a}$CCAST (World Laboratory), P.O.Box 8730, Beijing 100080,
China.}\\
{\small $^{b}$Department of Modern Physics, University of Science
        and Technology}\\
{\small of China (USTC), Hefei, Anhui 230027, China.}\\
{\small $^{c}$Institute of Theoretical Physics, Academia Sinica,} \\
{\small P.O.Box 2735, Beijing 100080, China.} }
\date{}
\maketitle

\vskip 12mm

\begin{center}\begin{minipage}{5in}

\vskip 5mm
\begin{center} {\bf Abstract}\end{center}
\baselineskip 0.3in
{The production of the lightest chargino pair from gluon-gluon
fusion is studied in the minimal supersymmetric model(MSSM) at
proton-proton colliders. We find that with the chosen parameters,
the production rate of the subprocess can be over 2.7 femto barn
when the chargino is higgsino-like, and the corresponding total
cross section in proton-proton collider can reach 56 femto barn
at the LHC in the CP-conserving MSSM. It shows that this loop
mediated subprocess can be competitive with the standard Drell-Yan
subprocess in proton-proton colliders, especially at the LHC.
Furthermore, our calculation shows it would be possible to extract
information about some CP-violating phase parameters, if we collected
enough chargino pair events. }\\

\vskip 5mm
{PACS number(s): 14.80.Ly, 12.15.Ji, 12.60.Jv}
\end{minipage}
\end{center}
\end{titlepage}

\baselineskip=0.36in

\eject
\rm
\baselineskip=0.36in

\begin{flushleft} {\bf 1. Introduction} \end{flushleft}
\par
The Minimal Standard Model(MSM) \cite{s1}\cite{s2} is a successful
theory of strong and electroweak interactions up to the present accessible
energies. Only the symmetric breaking sector of the electroweak
interactions remains to be directly tested by experiments. The
multi-TeV Large Hadron Collider LHC at CERN and the possible future
Next Linear Collider(NLC) are elaborately designed in order to
detect the symmetry-breaking mechanism and new physics beyond the MSM.
At present, the supersymmetric extended model(SUSY)\cite{haber}
\cite{Gunion} is widely considered as the theoretically most appealing
extension of the MSM. Apart from describing the experimental data as well
as the MSM does, the supersymmetric theory is able to solve various
theoretical problems, such as the fact that the SUSY model may provide an
elegant way to solve the deficiencies like the huge hierarchy problem,
the necessity of fine tuning and the non-occurrence of gauge coupling
unification at high energies.
\par
Searching for SUSY particles directly in experiment is one of the promising
tasks in the present and future colliders. The accurate measurements of the
sparticle production processes can give us much information about the
MSSM\cite{Han0}. Among various processes involving sparticles, chargino
pair production is one of the most important
reference processes of the MSSM which may appear firstly in $e^{+} e^{-}$,
$\gamma \gamma$ and hadron colliders.
The analyses treating chargino pair production at the theoretical level
are shown in references \cite{Diaz}\cite{Choi}\cite{Singo}\cite{Zhou1}
\cite{Zhou2}. So far no experimental evidence for charginos has been found
at LEP2,  and only the lower bound on the lightest chargino mass
$m_{\tilde{\chi}_{1}^{\pm}}$ is given. Recent experimental reports show
that the mass of lightest chargino may be larger than $90 GeV$\cite{L3}
\cite{OPAL}\cite{ALEPH}\cite{DELPHI}, and this bound depends mainly on
the sneutrino mass and the mass difference between the chargino and the
lightest SUSY particles(LSP) theoretically.
\par
The precise measurements of chargino pair production rates and chargino masses
give the possibilities of measuring some gaugino, higgsino couplings and
constraining the mass scale of squarks, which might not be in direct reach
in colliders. Several mechanisms can induce the production of chargino pair
at $pp$ colliders. One is through the Drell-Yan mechanism of
quark and antiquark, and another is by gluon-gluon fusion.
Although the chargino pair production via gluon-gluon fusion is a one-loop
process, the production rate can be significant due to the large
gluon luminosity in hadron colliders. In this paper we concentrate
on the capability of the lightest chargino pair production via gluon-gluon
collisions at $pp$ colliders in frame of the MSSM with full one-loop Feynman
diagrams. The paper is organized as follows: In section II, we introduce
some features of the model concerning in this work. In section III we
present the analytical expressions of the cross section for the subprocess
$gg \rightarrow \tilde{\chi}_{1}^{+} \tilde{\chi}_{1}^{-}$.
In section IV, we study the numerical results of the cross sections both for
subprocess and parent process. Finally, a short summary is presented.
In the Appendix, the relevant Feynman rules and some lengthy expressions
of the form factors appearing in the cross section in section III are listed.

\begin{flushleft} {\bf 2. The relevant theory of the MSSM.} \end{flushleft}
\vskip 5mm
\begin{center} {\bf A. The chargino-sector of the MSSM.} \end{center}
\par
In the MSSM theory the physical chargino mass eigenstates
$\tilde{\chi}^{\pm}_{1,2}$ are the combinations of the charged gauginos
and higgsinos. Their physical masses can be obtained by diagonalizing
the corresponding mass matrix $X$\cite{haber}. In the CP-noninvariant
MSSM theory, the mass term for charginos in lagrangian is
$$
-{\cal L}_{m}=\sum_{i}\bar{\tilde{\chi}}^{+}_{i}(U^{\ast}XV^{-1})_{ii}
          \tilde{\chi}^{+}_{i}
\eqno{(2.1)}
$$
$$
      X = \left(
      \begin{array}{ll}
         M_{SU(2)} & m_{W}\sqrt{2}\sin\beta  \\
         m_{W}\sqrt{2}\cos\beta &  |\mu|e^{i\phi_{\mu}}
      \end{array}
      \right),
\eqno{(2.2)}
$$
The complex phase of gaugino mass parameter $M_{SU(2)}$ can be rotated
away by field transformation, so we set $M_{SU(2)}$ to be real. $\mu$
is the higgsino mass parameter. The U, V are two $2 \times 2$ unitary
matrices defined to diagonalize the matrix $X$ to a diagonal matrix
$X_{D}$, namely,
$$
      U^{\ast}XV^{\dag} = X_{D},
\eqno{(2.3)}
$$
where $X_{D}$ has non-negative entries. The two diagonal elements of this
matrix can be expressed in a general form as\cite{Zhou1}\cite{Zhou}
\begin{eqnarray*}
      M_{\pm}^{2} &=& \frac{1}{2}
        \left\{
         M_{SU(2)}^{2}  + |\mu|^{2} + 2 m_{W}^{2} \pm
        \left[ (M_{SU(2)}^{2}-|\mu|^{2})^{2} + 4 m_{W}^{4} \cos^{2}2\beta +
        \right. \right. \\
&& \left. \left. 4 m_{W}^{2} (M_{SU(2)}^{2} +
         |\mu|^{2}+ 2 M_{SU(2)} |\mu|
         \sin2\beta \cos\phi_{\mu})
                     \right]^{1/2}
                \right\},
\hskip 20mm (2.4.1)
\end{eqnarray*}
which just stands for the expression of masses of charginos
$\tilde{\chi}^{\pm}_{1}$ and $\tilde{\chi}^{\pm}_{2}$. Inverting equation
(2.4.1), the fundamental SUSY parameters $M_{SU(2)}$ and $|\mu|$ can be
obtained from the alternative expressions on the right-hand side of the
following equations, respectively\cite{Zhou1}.
$$
(M_{SU(2)}, |\mu|)=\frac{1}{2} \left(
   \sqrt{m^2_{\tilde{\chi}^{+}_{1}}+m^2_{\tilde{\chi}^{+}_{2}}-2 m_W^2+
   2 M_{c}^2} \pm \sqrt{m^2_{\tilde{\chi}^{+}_{1}}+m^2_{\tilde{\chi}^{+}_{2}}-
   2 m_W^2-2 M_{c}^2} \right),
\hskip 5mm (2.4.2)
$$
where
$$
M_{c}^2= m_W^2 \cos{\phi_{\mu}} \sin{2 \beta}+\sqrt{m^2_{\tilde{\chi}^{+}_{1}} m^2_{\tilde{\chi}^{+}_{2}}-m_W^4
     \sin^2 2\beta \sin^2\phi_{\mu} }.\hskip 5mm (2.4.3)
$$
The diagonalizing matrices U and V are dependent on the complex phase of
$\mu$ and can be written in a general form as
$$
      U = \left(
      \begin{array}{ll}
         \cos\theta_{U}e^{i(\phi_{1}+\xi_{1})} &
         \sin\theta_{U}e^{i(\phi_{1}+\xi_{1}+\delta_{U})} \\
        -\sin\theta_{U}e^{i(\phi_{2}+\xi_{2}-\delta_{U})} &
         \cos\theta_{U}e^{i(\phi_{2}+\xi_{2})}
      \end{array}
      \right)
$$
$$
      V = \left(
      \begin{array}{ll}
         \cos\theta_{V}e^{i(\phi_{1}-\xi_{1})} &
         \sin\theta_{V}e^{i(\phi_{1}-\xi_{1}+\delta_{V})} \\
        -\sin\theta_{V}e^{i(\phi_{2}-\xi_{2}-\delta_{V})} &
         \cos\theta_{V}e^{i(\phi_{2}-\xi_{2})}
      \end{array}
      \right),
\eqno{(2.5)}
$$
In above equations the $\xi_{1}$ and $\xi_{2}$ are arbitrarily chosen phases.
It indicates the matrices U and V satisfying Eq.(2.3) are not unique, namely,
some arbitrary phases may be introduced into the physical fields. But our
analysis shows that they have no effects on the CP-odd observables.
The explicit forms of the other mixing and phase angles depending on the
Lagrangian parameters are given as\cite{Zhou}
$$
\tan{\theta_{U}}=\sqrt{\frac{M_{+}^{2}-M_{SU(2)}^{2}-2m_{W}^{2}\sin^{2}\beta}
                 {M_{+}^{2}-|\mu|^{2}-2m_{W}^{2}\cos^{2}\beta}},
$$
$$
\tan{\theta_{V}}=\sqrt{\frac{M_{+}^{2}-M_{SU(2)}^{2}-2m_{W}^{2}\cos^{2}\beta}
                 {M_{+}^{2}-|\mu|^{2}-2m_{W}^{2}\sin^{2}\beta}},
$$
$$
e^{i2\phi_{1}}=\frac{\cos\theta_{U}}{\cos\theta_{V}} \cdot
               \frac{M_{+}^{2}+M_{SU(2)}|\mu|\tan\beta e^{i\phi_{\mu}}
                              -2m_{W}^{2}\sin^{2}\beta}
               {M_{+}(M_{SU(2)}+|\mu| \tan\beta e^{i\phi_{\mu}})},
$$
$$
e^{i2\phi_{2}}=\frac{\cos\theta_{V}}{\cos\theta_{U}} \cdot
               \frac{M_{-}^{2}+M_{SU(2)}|\mu|\tan\beta e^{i\phi_{\mu}}
                              -2m_{W}^{2}\sin^{2}\beta}
               {M_{-}(M_{SU(2)}\tan\beta+|\mu| e^{-i\phi_{\mu}})},
$$
$$
e^{i\delta_{U}}=\frac{M_{SU(2)}+|\mu| e^{i\phi_{\mu}}\tan\beta}
                {|M_{SU(2)}+|\mu| e^{i\phi_{\mu}}\tan\beta|},
$$
$$
e^{i\delta_{V}}=\frac{M_{SU(2)}\tan\beta+|\mu| e^{i\phi_{\mu}}}
                {|M_{SU(2)}\tan\beta+|\mu| e^{i\phi_{\mu}}|},
\eqno{(2.6)}
$$

\vskip 5mm
\begin{center} {\bf B. The squark-sector of the MSSM.} \end{center}

\par
In the frame of the MSSM every quark has two scalar partners, the squarks
$\tilde{q}_L$ and $\tilde{q}_R$. If there is no left-right squark mixing
in the squark-sector, the mass matrix of a scalar quark including
CP-odd phases in the mass term of lagrangian, takes the following form\cite{s9}:
$$
 -{\cal L}_{m}=\left( \begin{array}{ll}
        \tilde{q}^{\ast}_{L} &  \tilde{q}^{\ast}_{R}
              \end{array} \right)
       \left( \begin{array}{ll}  m^2_{\tilde{q}_{L}} &  a_{q} m_{q} \\
           a_{q}^{\ast} m_{q}  &  m^2_{\tilde{q}_{R}}
              \end{array}  \right)
       \left( \begin{array}{ll}  \tilde{q}_{L}  \\  \tilde{q}_{R}
              \end{array}  \right),
\eqno{(2.7)}
$$
where $\tilde{q}_{L}$ and $\tilde{q}_{R}$ are the current eigenstates and
for the up-type scalar quarks, we have
$$m^2_{\tilde{q}_{L}}=\tilde{M}^2_{Q} + m^2_{q} +
      m_{Z}^2 (\frac{1}{2}- Q_q s_{W}^2) \cos{2 \beta},
\eqno{(2.8)} $$
$$m^2_{\tilde{q}_{R}}=\tilde{M}^2_{U} + m^2_{q} +
      Q_q m_{Z}^2 s_{W}^2 \cos{2 \beta}, \eqno{(2.9)} $$
$$
a_{q}=|a_{q}|e^{-2i \phi_{q}}=\mu \cot{\beta} + A^{\ast}_{q} \tilde{M}.
\eqno{(2.10)} $$
For the down-type scalar quarks,
$$m^2_{\tilde{q}_{L}}=\tilde{M}^2_{Q} + m^2_{q}
      - m_{Z}^2 (\frac{1}{2}+ Q_q s_{W}^2) \cos{2 \beta}, \eqno{(2.11)} $$
$$m^2_{\tilde{q}_{R}}=\tilde{M}^2_{D} + m^2_{q}
      + Q_q m_{Z}^2 s_{W}^2 \cos{2 \beta}, \eqno{(2.12)} $$
$$
a_{q}=|a_{q}|e^{-2i \phi_{q}}=\mu \tan{\beta} + A^{\ast}_{q} \tilde{M},
\eqno{(2.13)} $$
where $Q_{q}$($Q_{D}=-\frac{1}{3}$, $Q_{U}=\frac{2}{3}$) is the
charge of the scalar quark, $\tilde{M}^2_{Q}$, $\tilde{M}^2_{U}$ and
$\tilde{M}^2_{D}$ are the self-supersymmetry-breaking mass terms for
the left-handed and right-handed scalar quarks, $s_W=\sin{\theta_W}$,
$c_W=\sin{\theta_W}$. We choose $\tilde{M}_Q = \tilde{M}_U = \tilde{M}_D
= \tilde{M}$. $A_{q} \cdot \tilde{M}$ is a trilinear scalar interaction
parameter. The complex value $a_{q}$ can introduce CP-violation. In general,
$\tilde{q}_L$ and $\tilde{q}_R$ are mixed and give the mass eigenstates
$\tilde{q}_1$ and $\tilde{q}_2$(usually we assume $m_{\tilde{q}_1} <
m_{\tilde{q}_2}$). The mass eigenstates $\tilde{q}_1$ and $\tilde{q}_2$
are linear combinations of the current eigenstates $\tilde{q}_L$,
$\tilde{q}_R$
$$
\tilde{q}_1=\tilde{q}_L \cos{\theta_q} e^{ i\phi_{q}} -
            \tilde{q}_R \sin{\theta_q} e^{-i\phi_{q}},
$$
$$
\tilde{q}_2= \tilde{q}_L \sin{\theta_q} e^{ i\phi_{q}} +
              \tilde{q}_R \cos{\theta_q} e^{-i\phi_{q}},
\eqno{(2.14)}
$$
and
$$
\tan{2 \theta_q}=\frac{2 |a_{q}| m_{q}}
                      {m^2_{\tilde{q}_{L}}-m^2_{\tilde{q}_{R}} }.
\eqno{(2.15)}
$$
\par
where the $\theta_{q}$ is the mixing angle and $\phi_{q}$ is the CP-violating
phase. Then the masses of $\tilde{q}_1$ and $\tilde{q}_2$ are expressed as
$$
(m^2_{\tilde{q}_1},m^2_{\tilde{q}_2})=\frac{1}{2} \{
  m^2_{\tilde{q}_L} + m^2_{\tilde{q}_R} \mp  [ (m^2_{\tilde{q}_L} -
  m^2_{\tilde{q}_R})^2 + 4 |a_q|^2 m_q^2 ] ^{\frac{1}{2}} \}.
\eqno{(2.16)}
$$
In the CP-violating MSSM theory, there are several possibilities to introduce
CP-odd phases\cite{dim}. In our process, two CP-odd phases are
involved, respectively appearing in the squark mass and chargino mass matrices.
The detailed analyses of the present upper bounds on electron and neutron
electric dipole moments may give constraints on CP-odd phase parameters
indirectly\cite{Ibr}. But these constraints should be rather
weak, since those results depend strongly on the assumptions which are used.
Recently S.Y. Choi et al discussed the impacts of the CP-odd phase stemming
from chargino mass matrix in the production of lightest chargino-pair in
$e^+ e^-$ collisions at the tree-level\cite{Achoi}. In reference \cite{Zhou}
\cite{Zhou1} the effects from the CP-odd phases in the processes of the top
quark pair and lightest chargino pair productions in $\gamma \gamma$
collisions at one-loop level in frame of the CP-violating MSSM, are
investigated. There all the relevant CP-odd complex phases are kept and
the CP-violating effects are studied without any extra limitations on
CP-odd phases for the general discussion. In this work we shall investigate
the CP-odd effects in the same way as in previous works\cite{Zhou}
\cite{Zhou1}.

\vskip 5mm
\begin{center} {\bf C. The Higgs-sector of the MSSM.} \end{center}
\par
The supersymmetric model requires at least the extension of one additional
Higgs-doublet where the parameters of the Higgs sector are tightly related.
In the MSSM the neutral Higgs boson masses $m_{h^0}$, $m_{H^0}$ and $m_{A^0}$
are given by\cite{Esp}
\begin{eqnarray*}
m_{h^0,H^0}^2 &=& \frac{1}{2}\left[ m_{A^0}^2+m_{Z}^2+\epsilon \mp  \right. \\
  && \left. \sqrt{(m_{A^0}^2+m_{Z}^2+\epsilon)^2-4m_{A^0}^2m_{Z}^2
  cos^2{2\beta}- 4\epsilon(m_{A^0}^2sin^2{\beta}+m_{Z}^2cos^2{\beta}) } \right],
\hskip 10mm (2.17)
\end{eqnarray*}
$$
m_{H^{\pm}}^2= m_{A^0}^2+m_{W}^2
\eqno{(2.18)}
$$
with the leading corrections being characterized by the radiative parameter
$\epsilon$
$$
\epsilon = \frac{3 G_{F}}{\sqrt{2}\pi^2} \frac{m_t^4}{sin^2{\beta}} log \left[
         \frac{m_{\tilde{t}}^2}{m_{t}^2} \right] .
\eqno{(2.19)}
$$
The parameter $m_{\tilde{t}}^2=m_{\tilde{t}_1}m_{\tilde{t}_2}$ denotes
the average squared mass of the stop quarks. The mixing angle $\alpha$
is fixed by $\tan{\beta}$ and the Higgs boson mass $m_{A^0}$,
\begin{eqnarray*}
\tan{2 \alpha}=\tan{2 \beta} \frac{m_{A^0}^2+m_{Z}^2} {m_{A^0}^2-m_{Z}^2
               + \frac{\epsilon}{\cos{2 \beta}} }  (-\frac{\pi}{2}<\alpha<0).
\hskip 10mm (2.20)
\end{eqnarray*}

\par
\begin{flushleft} {\bf 3. The calculation of subprocess $gg \rightarrow
                   \tilde{\chi}_1^{+} \tilde{\chi}_1^{-}$ in the MSSM.}
\end{flushleft}
\par
The process producing the lightest chargino pair via gluon-gluon collisions
can only be induced through one-loop diagrams. In this case
it is not necessary to consider the renormalization at one-loop level
and the ultraviolet divergence should be canceled automatically
if all the one-loop diagrams in MSSM are included.
In this work, we perform the evolution in the t'Hooft-Feynman gauge.
The generic Feynman diagrams contributing to
the subprocess $gg \rightarrow \tilde{\chi}_1^{+} \tilde{\chi}_1^{-}$ in
frame of the MSSM are depicted in figure 1, where the diagrams with
exchanging incoming gluons are not shown except for quartic interaction
diagrams shown in Fig.(b.1 $ \sim $ 2). The relevant Feynman rules
can be found in Appendix A. All the one-loop diagrams can be divided into
three groups: (1) box diagrams shown in Fig.1(a.1 $ \sim $ 3). (2) quartic
interaction diagrams in Fig.1(b.1 $ \sim $ 2). (3) triangle diagrams shown
in Fig.1(c.1 $ \sim $ 2). In our calculation we find the contributions from
the $\gamma$ (or $Z^0$) exchanging s-channel Feynman diagrams with
quark loops as shown in Fig.1(c.2), are zero (or very small).
It can be understood by the Furry theorem. The Furry theorem forbids
the production of the spin-one components of the $Z^{0}$ and $\gamma$ from
a fermion loop. And the contribution from the spin-zero component of the
$Z^{0}$ vector boson coupling with a pair of charginos is very small.
The contribution from each $\gamma$ ($Z^{0}$) exchanging s-channel
diagram involving a squark loop shown in Fig.1(b.2) and Fig.1(c.1),
is canceled out by the same type diagram, but involving its corresponding
anti-squark loop. Here we should mention that there are also some diagrams
having no contribution to the process, such as the s-channel diagrams
with trilinear gluon interactions. Since the vertices of
$A^{0}(G^{0})-\tilde{q}_{i}-\tilde{q}_{i}$ vanish\cite{Gunion},
there is no such diagram with a triangle squark loop coupling with
$A^{0}$ or $G^{0}$ Higgs boson. All these diagrams are not drawn in Fig.1.
\par
  In this work, we denote the reaction of chargino-pair production
via gluon-gluon collisions as:
$$
g (p_3, \mu) g (p_4, \nu) \longrightarrow
            \tilde{\chi}_{1}^{+} (p_1) \tilde{\chi}_{1}^{-} (p_2).
$$
We write the corresponding matrix element for each of the diagrams
in Fig.1 in the form according to their Lorentz invariant structure:
\begin{eqnarray*}
\delta {\cal M} &=&
    {\cal M}^{b}+{\cal M}^{tr}+{\cal M}^{q} \\
&=& {\cal M}^{b,\hat{t}}+{\cal M}^{b,\hat{u}}+{\cal M}^{tr,\hat{t}}+
    {\cal M}^{tr,\hat{u}}+{\cal M}^{q} \\
&=& \frac{1}{2} \delta_{ab} \epsilon^{\mu}(p_3)\epsilon^{\nu}(p_4)
    \bar{u}(p_1) \left\{
    f_{1} \gamma_{\mu}\gamma_{\nu}+ f_{2} \gamma_{\nu}\gamma_{\mu}+
    f_{3} \gamma_{\mu}p_{1\nu}+ f_{4} \gamma_{\mu}p_{2\nu} \right. \\
&& \left. + f_{5} \gamma_{\nu}p_{1\mu}+ f_{6} \gamma_{\nu}p_{2\mu}+
    f_{7} p_{1\mu}p_{1\nu}+ f_{8} p_{1\mu}p_{2\nu}+
    f_{9} p_{1\nu}p_{2\mu}+ f_{10} p_{2\mu}p_{2\nu} \right. \\
&& \left. + f_{11} \rlap/{p}_{3}\gamma_{\mu}\gamma_{\nu}+
    f_{12} \rlap/{p}_{3}\gamma_{\nu}\gamma_{\mu}+
    f_{13} \rlap/{p}_{3}\gamma_{\mu}p_{1\nu}+
    f_{14} \rlap/{p}_{3}\gamma_{\mu}p_{2\nu} \right. \\
&& \left. + f_{15} \rlap/{p}_{3}\gamma_{\nu}p_{1\mu}+
    f_{16} \rlap/{p}_{3}\gamma_{\nu}p_{2\mu}+
    f_{17} \rlap/{p}_{3}p_{1\mu}p_{1\nu}+
    f_{18} \rlap/{p}_{3}p_{1\mu}p_{2\nu} \right. \\
&& \left. + f_{19} \rlap/{p}_{3}p_{1\nu}p_{2\mu}+
    f_{20} \rlap/{p}_{3}p_{2\mu}p_{2\nu}+
    f_{21} \epsilon_{\mu\nu\alpha\beta} p_{1}^{\alpha} p_{3}^{\beta}+
    f_{22} \epsilon_{\mu\nu\alpha\beta} p_{2}^{\alpha} p_{3}^{\beta} \right. \\
&& \left. + f_{23} \epsilon_{\mu\nu\alpha\gamma} p_{1}^{\alpha} \gamma^{\gamma}+
    f_{24} \epsilon_{\mu\nu\alpha\gamma} p_{2}^{\alpha} \gamma^{\gamma}+
    f_{25} \epsilon_{\mu\nu\alpha\gamma} p_{3}^{\alpha} \gamma^{\gamma}+
    f_{26} \epsilon_{\mu\alpha\beta\gamma} p_{1}^{\alpha} p_{3}^{\beta}
    \gamma^{\gamma} p_{1\nu} \right. \\
&& \left. + f_{27} \epsilon_{\mu\alpha\beta\gamma} p_{1}^{\alpha} p_{3}^{\beta}
    \gamma^{\gamma} p_{2\nu}+
    f_{28} \epsilon_{\mu\alpha\beta\gamma} p_{2}^{\alpha} p_{3}^{\beta}
    \gamma^{\gamma} p_{1\nu}+
    f_{29} \epsilon_{\mu\alpha\beta\gamma} p_{2}^{\alpha} p_{3}^{\beta}
    \gamma^{\gamma} p_{2\nu} \right. \\
&& \left. + f_{30} \epsilon_{\nu\alpha\beta\gamma} p_{1}^{\alpha}
    p_{3}^{\beta} \gamma^{\gamma} p_{1\mu}+
    f_{31} \epsilon_{\nu\alpha\beta\gamma} p_{1}^{\alpha} p_{3}^{\beta}
    \gamma^{\gamma} p_{2\mu}+
    f_{32} \epsilon_{\nu\alpha\beta\gamma} p_{2}^{\alpha} p_{3}^{\beta}
    \gamma^{\gamma} p_{1\mu} \right. \\
&& \left. + f_{33} \epsilon_{\nu\alpha\beta\gamma} p_{2}^{\alpha} p_{3}^{\beta}
    \gamma^{\gamma} p_{2\mu}+
    f_{34} \gamma_{5}\gamma_{\mu}\gamma_{\nu} +
    f_{35} \gamma_{5}\gamma_{\nu}\gamma_{\mu} \right. \\
&& \left. + f_{36} \gamma_{5}\gamma_{\mu}p_{1\nu}+
    f_{37} \gamma_{5}\gamma_{\mu}p_{2\nu}+f_{38} \gamma_{5}\gamma_{\nu}p_{1\mu}+
    f_{39} \gamma_{5}\gamma_{\nu}p_{2\mu} \right. \\
&& \left. + f_{40} \gamma_{5}p_{1\mu}p_{1\nu}+ f_{41} \gamma_{5}p_{1\mu}p_{2\nu}+
    f_{42} \gamma_{5}p_{1\nu}p_{2\mu}+ f_{43} \gamma_{5}p_{2\mu}p_{2\nu}+
    f_{44} \gamma_{5}\rlap/{p}_{3}\gamma_{\mu}\gamma_{\nu} \right. \\
&& \left. + f_{45} \gamma_{5}\rlap/{p}_{3}\gamma_{\nu}\gamma_{\mu}+
    f_{46} \gamma_{5}\rlap/{p}_{3}\gamma_{\mu}p_{1\nu}+
    f_{47} \gamma_{5}\rlap/{p}_{3}\gamma_{\mu}p_{2\nu}+
    f_{48} \gamma_{5}\rlap/{p}_{3}\gamma_{\nu}p_{1\mu} \right. \\
&& \left. +f_{49} \gamma_{5}\rlap/{p}_{3}\gamma_{\nu}p_{2\mu}+
    f_{50} \gamma_{5}\rlap/{p}_{3}p_{1\mu}p_{1\nu}+
    f_{51} \gamma_{5}\rlap/{p}_{3}p_{1\mu}p_{2\nu}+
    f_{52} \gamma_{5}\rlap/{p}_{3}p_{1\nu}p_{2\mu} \right. \\
&& \left. + f_{53} \gamma_{5}\rlap/{p}_{3}p_{2\mu}p_{2\nu}+
    f_{54} \epsilon_{\mu\nu\alpha\beta} p_{1}^{\alpha} p_{3}^{\beta} \gamma_{5}+
    f_{55} \epsilon_{\mu\nu\alpha\beta} p_{2}^{\alpha} p_{3}^{\beta} \gamma_{5}+
    f_{56} \epsilon_{\mu\nu\alpha\gamma} p_{1}^{\alpha} \gamma_{5} \gamma^{\gamma} \right. \\
&& \left. + f_{57} \epsilon_{\mu\nu\alpha\gamma} p_{2}^{\alpha} \gamma_{5} \gamma^{\gamma} +
    f_{58} \epsilon_{\mu\nu\alpha\gamma} p_{3}^{\alpha} \gamma_{5}
    \gamma^{\gamma}+
    f_{59} \epsilon_{\mu\delta\alpha\gamma} p_{1}^{\delta} p_{3}^{\alpha}
    \gamma_{5} \gamma^{\gamma} p_{1\nu}+
    f_{60} \epsilon_{\mu\delta\alpha\gamma} p_{1}^{\delta} p_{3}^{\alpha}
    \gamma_{5} \gamma^{\gamma} p_{2\nu} \right. \\
&& \left. + f_{61} \epsilon_{\mu\delta\alpha\gamma} p_{2}^{\delta}
    p_{3}^{\alpha} \gamma_{5} \gamma^{\gamma} p_{1\nu}+
    f_{62} \epsilon_{\mu\delta\alpha\gamma} p_{2}^{\delta} p_{3}^{\alpha}
    \gamma_{5} \gamma^{\gamma} p_{2\nu}+
    f_{63} \epsilon_{\nu\delta\alpha\gamma} p_{1}^{\delta} p_{3}^{\alpha}
    \gamma_{5} \gamma^{\gamma} p_{1\mu} \right. \\
&& \left. + f_{64} \epsilon_{\nu\delta\alpha\gamma} p_{1}^{\delta}
    p_{3}^{\alpha} \gamma_{5} \gamma^{\gamma} p_{2\mu}+
    f_{65} \epsilon_{\nu\delta\alpha\gamma} p_{2}^{\delta} p_{3}^{\alpha}
    \gamma_{5} \gamma^{\gamma} p_{1\mu}+
    f_{66} \epsilon_{\nu\delta\alpha\gamma} p_{2}^{\delta} p_{3}^{\alpha}
    \gamma_{5} \gamma^{\gamma} p_{2\mu} \right\}v(p_{2}),
\hskip 10mm (3.1)
\end{eqnarray*}
with
$$
f_i = f_{i}^{b}+f_{i}^{tr}+f_{i}^{q} \hskip 20mm (i=1 \sim 66),
\eqno{(3.2)}
$$
where $\frac{1}{2}\delta_{ab}$ is the color factors in
amplitudes, ${\cal M}^{b}$, ${\cal M}^{tr}$ and ${\cal M}^{q}$ are
the matrix elements contributed by box, triangle and quartic interactions
diagrams, respectively. $f_{i}^{b}$, $f_{i}^{tr}$ and $f_{i}^{q}$ are their
corresponding form factors. As we divided the matrix elements
${\cal M}^{b}$ and ${\cal M}^{tr}$ into t- and u-channel parts, respectively,
so for each of the corresponding form factors we have
$$
f_{i}^{b}=f_{i}^{b,\hat{t}}+f_{i}^{b,\hat{u}}, \hskip 5mm
f_{i}^{tr}=f_{i}^{tr,\hat{t}}+f_{i}^{tr,\hat{u}} \hskip 5mm (i=1 \sim 66).
$$
Since the amplitude parts from the u-channel box and triangle vertex
interaction diagrams can be obtained from the t-channel's by doing
exchanges shown as below:
\begin{eqnarray*}
{\cal M}^{j,\hat{u}}={\cal M}^{j,\hat{t}}(\hat{t} \rightarrow \hat{u},
   p_3 \leftrightarrow p_4, \mu \leftrightarrow \nu),\hskip 3mm (j=b,tr)
\end{eqnarray*}
Then only the explicit t-channel form factors $f_i^{b,\hat{t}}$
and $f_i^{tr,\hat{t}} (i=1\sim 66)$ for box and triangle diagrams,
and the form factors for quartic interaction diagrams are listed
in Appendix B. The cross section for this subprocess at the one-loop order
in unpolarized photon collisions can be obtained by
\begin{eqnarray*}
 \hat{\sigma}(\hat{s}) &=& \frac{1}{16 \pi \hat{s}^2}
             \int_{\hat{t}^{-}}^{\hat{t}^{+}} d\hat{t}
         \bar{\sum\limits_{}^{}} |{\cal M}|^2, \hskip 10mm (3.3)
\end{eqnarray*}
where $\hat{t}^\pm=(m_{\tilde{\chi}^{+}_1}^2-\frac{1}{2}\hat{s})\pm
\frac{1}{2}\hat{s}\beta$. The bar over sum notation means that
we are doing average over initial spins and colors.

\par
\begin{flushleft} {\bf 4. Numerical results and discussions} \end{flushleft}
\par
In this section, we present some numerical results of the total cross
section from the full one-loop diagrams involving virtual (s)quarks for
the subprocess of $gg \rightarrow \tilde{\chi}^{+}_{1} \tilde{\chi}^{-}_{1}$
and parent process $pp \rightarrow gg \rightarrow \tilde{\chi}^{+}_{1}
\tilde{\chi}^{-}_{1}+X$, respectively. The general input parameters
involved are chosen as: $m_t=175 GeV$, $m_{Z}=91.187 GeV$, $m_b=4.5 GeV$,
$\sin^2{\theta_{W}}=0.2315$, and $\alpha = 1/137.036$.
We adopt a simple one-loop formula for the running strong coupling
constant $\alpha_s$ as
$$
\alpha_s(\mu)=\frac{\alpha_{s}(m_Z)} {1+\frac{33-2 n_f} {6 \pi} \alpha_{s}
              (m_Z) \ln \left( \frac{\mu}{m_Z} \right) }. \eqno{(4.1)}
$$
where $\alpha_s(m_Z)=0.117$ and $n_f$ is the number of active flavors at
energy scale $\mu$.
\par
We take the numerical values of the MSSM parameters in CP-conserving case
which are also acceptable in the frame of the Minimal Supergravity (mSUGRA)
model, since the mSUGRA is the simplest and most fully investigated model.
It assumes that the boundary conditions are set at $M_{U}$. The ranges of
the model parameters should be constrained by the evolution to low energies
${}^{<}_{\sim} m_{SUSY}$ beginning with the boundary conditions at $M_{U}$.
With this consideration we take the following parameter values by default
unless otherwise stated.
\par
The squark masses of the first two generations are approximately degenerated,
namely, we can neglect their mixing angles between the left- and right-squarks.
and choose $m_{\tilde{u}_{1,2}}=m_{\tilde{d}_{1,2}}=m_{\tilde{c}_{1,2}}
=m_{\tilde{s}_{1,2}}=600 GeV$
From renormalization group equations \cite{Drees2} one expects that
the soft SUSY breaking masses $m_{\tilde{q}_{L}}$ and $m_{\tilde{q}_{R}}$
of the third generation squarks are smaller than those of the first and
second generations due to the Yukawa interactions. The third family stop
quarks are normally significantly mixed and split due to the large mass
of the top quark, and the lightest scalar top quark mass eigenstate
$\tilde{t}_1$ can be much lighter than the top quark and all the scalar
partners of the light quarks. Therefore we assume
$m_{\tilde{t}_1} < m_{\tilde{t}_2}$ and $\theta_{t} \sim \frac{\pi}{4}$,
and take $\tilde{M}_Q = \tilde{M}_U = \tilde{M}_D = \tilde{M} = 200 GeV$
for the third generation squarks. For simplicity, we set the sbottom
mixing angle being zero ($\theta_{b}=0$). Then the masses of stop, sbottom
mass eigenstates $\tilde{t}_{1,2}$ and $\tilde{b}_{1,2}$ can be determined
quantitatively by Eqs.(2.8 $ \sim $ 2.16). For the CP-odd neutral Higgs mass
$m_{A^0}$ we set its value being typical large and take $m_{A^0}=250 GeV$.
The ratio of the vacuum expectation values $\tan{\beta}$ is chosen to be 4
or 40 in order to make comparison. The masses of other Higgs bosons can be
obtained from Eqs.(2.17) $\sim$ (2.20). We checked that with these
input parameters the experimental limits on the masses of Higgs bosons
are not violated. Since these mass values of Higgs bosons are far below
the threshold of chargino pair production in our numerical calculation,
we set all the decay widths of Higgs bosons to be 10 GeV and these value
choices will not influence our results significantly. In the CP-violating
case, we use also the above input parameters for comparison.
\par
The physical chargino masses $m_{\tilde{\chi}^{+}_{1}}$ and
$m_{\tilde{\chi}^{+}_{2}}$ are taken to be 165 GeV and 750 GeV, respectively.
The fundamental SUSY parameters $M_{SU(2)}$ and $|\mu|$ can be extracted at
the tree level from these input chargino masses, $\tan{\beta}$ and the complex
phase angle of $\mu$ by using Eqs.(2.4.2 $ \sim $ 3). When $\mu$ is real,
we assume $\mu$ is positive. Then the lightest chargino is dominantly gaugino
(gaugino-like or wino-like) when there is $M_{SU(2)} << \mu$, and the
chargino is dominantly higgsino (higgsino-like), when $M_{SU(2)}$ is much
larger than $\mu$. In the following we shall investigate the numerical
results in both extreme cases.
\par
The total cross sections of the lightest chargino pair production
via gluon fusion as the functions of the c.m.s. energy of gluons
$\sqrt{\hat{s}}$ with $m_{\tilde{\chi}^{+}_{1}}=165 GeV$,
$m_{\tilde{\chi}^{+}_{2}}=750 GeV$ and all vanishing CP phases, are shown
in Fig.2(a) and (b). In figure 2(a) the two curves correspond to the
higgsino-like chargino case with $\tan{\beta}=4$ and
$\tan{\beta}=40$, respectively. Whereas the plot in Figure 2(b)
is for gaugino-like chargino case. It is obvious that the subprocess cross
section of the pair production of the lightest higgsino-like chargino is
one order larger than that of the gaugino-like chargino pair production.
And in general, the cross sections with $\tan{\beta}=40$ are approximately
one to four times larger than those with $\tan{\beta}=4$. Because of the
resonance effects, all the four curves in Fig.2(a) $\sim$ (b) have peaks
and spikes at the energy positions where the resonance conditions are
satisfied. There are turn points on all four curves, which are located
at the vicinity of $\sqrt{\hat{s}}=2 m_{t}=350 GeV$. On the two curves
of Fig.2(a,b) with $\tan{\beta}=4$, there are two small spikes stemming
from resonance effects in the vicinities of $\sqrt{\hat{s}} \sim
2 m_{\tilde{b}_{1}} \sim 403 GeV$ and $\sqrt{\hat{s}} \sim 2 m_{\tilde{b}_{2}}
\sim 415 GeV$, respectively. Whereas for the other two
curves in Fig.2(a,b) with $\tan{\beta}=40$, the small spikes due
to resonance effect are located at the positions of $\sqrt{\hat{s}} \sim
2 m_{\tilde{b}_{1}} \sim 403 GeV$ and $\sqrt{\hat{s}}=2 m_{\tilde{b}_2}
\sim 417 GeV$, respectively.
\par
Figure 3 gives the total cross sections of the subprocess as the functions
of the lightest higgsino-like chargino mass with $\sqrt{\hat{s}}=450 GeV$.
In this figure we can see considerable enhancement around the
region of $m_{\tilde{\chi}^{+}_1}=185 GeV$.
Figure 4 gives the cross sections of the higgsino-like chargino pair production
subprocess as a function of self-supersymmetry-breaking mass parameter of the
third generation scalar quarks $\tilde{M}$, when $\sqrt{\hat{s}}=450 GeV$
(Here we set the masses for squarks of the first and second generations,
are degenerated and have the values being $600 GeV$.). The two curves have
obvious spikes due to resonance effect at the positions of $\tilde{M}=215 GeV$
for both $\tan{\beta}=4$ and $\tan{\beta}=40$, respectively. There we have
$\sqrt{\hat{s}}=450 GeV \sim  2 m_{\tilde{b}_{1,2}}$.
\par
The cross sections in the subprocess of higgsino-like
chargino pair production versus the CP phases angles $\phi_{CP}(=\phi_{\mu},
\phi_{q})$ (Here we take $\phi_{q}=\phi_{t}=\phi_{b}$ and $\phi_{u,d,c,s}=0$)
with $\sqrt{\hat{s}}=450 GeV$, $m_{\tilde{\chi}^{+}_1}=165 GeV$ and
$m_{\tilde{\chi}^{+}_2}=750 GeV$, are depicted in figure 5(a) and 5(b).
Fig.5(a) is for $\tan{\beta}=40$ and Fig.5(b) for $\tan{\beta}=4$.
In both figures, the full-lines and dotted-lines correspond to
$\phi_{CP}=\phi_q(q=t,b)$ and $\phi_{CP}=\phi_{\mu}$, respectively.
The curves in Fig.5(a,b) show the periodical features of
$\hat{\sigma}(\phi_q)=\hat{\sigma}(\pi+\phi_q)$ for the curves of
$\hat{\sigma}$ versus $\phi_q$ and $\hat{\sigma}(\phi_{\mu})=
\hat{\sigma}(2 \pi+\phi_{\mu})$ for the curves of $\hat{\sigma}$
versus $\phi_{\mu}$, respectively. All the two CP phase
angles affect the cross sections obviously, but the effects from
the phase angle $\phi_{q}$ are a little stronger than those from $\phi_{\mu}$.
\par
With the chargino pair production rate in gluon-gluon fusion, we can easily
obtain the total cross section in $pp$ collider, by folding the cross
section of the subprocess $\hat{\sigma} (gg \rightarrow
\tilde{\chi}^{+}_1 {\tilde{\chi}}^{-}_1)$ with the gluon luminosity.
$$
\sigma(pp \rightarrow gg \rightarrow
          \tilde{\chi}^{+}_1 \tilde{\chi}^{-}_1+X)=
    \int_{4 m_{\tilde{\chi}^{+}_{1}}^2/s }^{1}
d\tau \frac{dL_{gg}}{d\tau} \hat{\sigma}(gg \rightarrow
          \tilde{\chi}^{+}_1 \tilde{\chi}^{-}_1,
     \hskip 3mm at \hskip 3mm \hat{s}=\tau s). \eqno{(4.2)}
$$
where $\sqrt{s}$ and $\sqrt{\hat{s}}$  denote the proton-proton
and gluon-gluon c.m.s. energies respectively and
$\frac{dL_{gg}} {d\tau}$ is the gluon luminosity, which is defined as
$$
\frac{d{\cal L}_{gg}}{d\tau}=\int_{\tau}^{1}
 \frac{dx_1}{x_1} \left[ F_{g}(x_1,\mu) F_{g}(\frac{\tau}{x_1},\mu) \right].
 \eqno{(4.3)}
$$
Here we used $\tau=x_{1}x_{2}$, one can find the definitions of
$x_1$ and $x_2$ in Ref.\cite{s6} and the energy scale $\mu$ is taken as
$\mu=\sqrt{\hat{s}}$. We adopt the MRS set G parton distribution function
$F_{g}(x)$ \cite{ss1}, and ignore the supersymmetric QCD corrections
to the parton distribution functions for simplicity. The numerical calculation
is carried out for the LHC at the energy around $10 \sim 14 TeV$.
\par
The cross section for the process of the lightest higgsino-like chargino
pair production $pp \rightarrow gg \rightarrow \tilde{\chi}^{+}_1
\tilde{\chi}^{-}_1+X$ versus $\sqrt{s}$, with $m_{\tilde{\chi}^{+}_{1}}=
165 GeV$, $m_{\tilde{\chi}^{+}_{2}}=750 GeV$, are depicted in Fig.6.
The full and dashed lines are for $\tan{\beta}=4$ and 40 respectively,
with all CP phase angles being zero. The dotted-line is for
$\phi_{t}=\phi_{b} =\pi/4$ and other CP phase angels being vanished.
We can see that the total cross section at the future
LHC collider can reach 56 femto barn for the higgsino-like chargino pair
production, when $\sqrt{s} \sim 14 TeV$, $\tan{\beta}=40$ and all CP phase
angles vanish. Calculating with the analytical expressions given in
Ref.\cite{Hong}, the results show that with the same input parameters, when
$\sqrt{s} \sim 14 TeV$ and $\tan{\beta}=40$, the cross section of the lightest
chargino pair production via quark-antiquark annihilation can reach 317 femto
barn for higgsino-like case in the CP-conserving MSSM theory. Therefore we can
conclude that the chargino pair production via gluon-gluon fusion is
competitive with the standard Drell-Yan production at the LHC and can be
considered as a part of the NLO QCD correction to the Drell-Yan production
process. In Fig.6 we can see that the production rate has the weak dependence
on the c.m.s energy $\sqrt{s}$ for $\tan{\beta}=4$, but is strongly related
to $\sqrt{s}$ for $\tan{\beta}=40$. The CP-violating effect in total cross
section of the lightest chargino pair production in the LHC, is also obvious.
The discrepancies between the total cross sections of the lightest chargino
pair production predicted in the CP-conserving and the CP-violating MSSM at
the LHC, are about $20\%$ as shown in Fig.6.

\par
\begin{flushleft} {\bf 5. Summary} \end{flushleft}
\par
In this paper, we studied the pair production process of the lightest
chargino via gluon-gluon fusion at the LHC. The numerical
analyse of its production rates is carried out in the MSSM with typical
parameter sets. The results show that the cross section of
the lightest chargino pair production via gluon-gluon fusion can be
over 2.7 femto barn and the cross section at a future LHC collider can be 6.2
to 56 femto barn for the higgsino-like chargino pair production.
It shows clearly that the production rates in proton-proton colliders can
be largely enhanced if the chargino is higgsino-like. We find that the
chargino production via gluon-gluon fusion could be competitive
with the standard Drell-Yan production in the LHC and can be considered
as a part of the NLO QCD correction to the Drell-Yan production subprocess.
Our calculation shows also that in some exceptional c.m.s energy regions
of incoming gluons, where the resonance conditions are satisfied in the
parameter space, we can see observable enhancement effects on the curves.
We also investigated the effects of complex phases $\phi_{q}$ in the squark
mass matrices and $\phi_{\mu}$ appearing in chargino mass matrix in
higgsino-like chargino case and found that the production rates in subprocess
are sensitive to the CP-odd complex phases $\phi_{q}$ and $\phi_{\mu}$,
but the effects from the phase angles $\phi_{q}$ are stronger than those from
$\phi_{\mu}$. The effects from the CP-odd phase angles can be also
demonstrated in the total cross section of the lightest chargino pair
production in the LHC. Therefore it could be possible to get some
information about these phase parameters, if we collected enough events
statistically in searching for chargino pair via gluon-gluon fusion at
the LHC.

\vskip 4mm
\noindent{\large\bf Acknowledgement:}
These work was supported in part by the National Natural Science
Foundation of China(project numbers: 19675033, 19875049), the Youth Science
Foundation of the University of Science and Technology of China and
a grant from the Education Ministry of China.

\vskip 5mm
\begin{center} {\Large Appendix}\end{center}
\par
A. The relevant Feynman rules of the MSSM.
\par
  The Feynman rules for the couplings of $q-\tilde{q}^{\prime}_{L,R}-
\tilde{\chi}^{+}_{j=1,2}$ are presented in Ref.\cite{haber}\cite{Gunion}. Then
the corresponding Feynman rules for such vertices in squark mass eigenstate
basis can be obtained as:
$$
\bar{U}-\tilde{D}_{i}-\tilde{\chi}^{+}_{j}: \hskip 5mm
V_{U\tilde{D}_{i}\tilde{\chi}^{+}_{j}}^{(1)}P_L+
V_{U\tilde{D}_{i}\tilde{\chi}^{+}_{j}}^{(2)}P_R,
\eqno{(A.1.1)}
$$
$$
U-\bar{\tilde{D}}_{i}-\bar{\tilde{\chi}}^{+}_{j}:\hskip 5mm
-V_{U\tilde{D}_{i}\tilde{\chi}^{+}_{j}}^{(2)\ast}P_L-
V_{U\tilde{D}_{i}\tilde{\chi}^{+}_{j}}^{(1)\ast}P_R,
\eqno{(A.1.2)}
$$

$$
D-\bar{\tilde{U}}_{i}-\bar{\tilde{\chi}}^{+c}_{j}:\hskip 5mm
C^{-1}\left\{ V_{D\tilde{U}_{i}\tilde{\chi}^{+}_{j}}^{(1)}P_L+
V_{D\tilde{U}_{i}\tilde{\chi}^{+}_{j}}^{(2)}P_R \right\},
\eqno{(A.1.3)}
$$
$$
\bar{D}-\tilde{U}_{i}-\tilde{\chi}^{+c}_{j}:\hskip 5mm
\left\{ V_{D\tilde{U}_{i}\tilde{\chi}^{+}_{j}}^{(2)\ast}P_L+
V_{D\tilde{U}_{i}\tilde{\chi}^{+}_{j}}^{(1)\ast}P_R \right\} C,
\eqno{(A.1.4)}
$$
respectively. Here $(U,D)=(u,d),(c,s),(t,b)$ and C is the charge conjugation
matrix, which appears when there is a discontinuous flow of fermion number,
$P_{L,R}=\frac{1}{2} (1\mp\gamma_5)$ and
$$
V_{U\tilde{D}_{1}\tilde{\chi}^{+}_{j}}^{(1)}= \frac{i g m_U}
  {\sqrt{2}m_W \sin{\beta}} V_{j2}^{\ast} \cos{\theta_{D}} e^{-i\phi_D},
\eqno{(A.2.1)}
$$
$$
V_{U\tilde{D}_{1}\tilde{\chi}^{+}_{j}}^{(2)}= -i g (U_{j1} \cos{\theta_D}
   e^{-i\phi_D} + \frac{m_D}{\sqrt{2}m_W \cos{\beta}} U_{j2} \sin{\theta_D}
   e^{i\phi_D}),
\eqno{(A.2.2)}
$$
$$
V_{U\tilde{D}_{2}\tilde{\chi}^{+}_{j}}^{(1)}= \frac{i g m_U}
  {\sqrt{2}m_W \sin{\beta}} V_{j2}^{\ast} \sin{\theta_{D}} e^{-i\phi_D},
\eqno{(A.2.3)}
$$
$$
V_{U\tilde{D}_{2}\tilde{\chi}^{+}_{j}}^{(2)}= -i g (U_{j1} \sin{\theta_D}
   e^{-i\phi_D}- \frac{m_D}{\sqrt{2}m_W \cos{\beta}} U_{j2} \cos{\theta_D}
   e^{i\phi_D}),
\eqno{(A.2.4)}
$$
$$
V_{D\tilde{U}_{1}\tilde{\chi}^{+}_{j}}^{(1)}= ig (V_{j1}^{\ast} \cos{\theta_U}
   e^{i\phi_U} + \frac{m_U}{\sqrt{2}m_W \sin{\beta}} V_{j2}^{\ast}
   \sin{\theta_U} e^{-i\phi_D}),
\eqno{(A.3.1)}
$$
$$
V_{D\tilde{U}_{1}\tilde{\chi}^{+}_{j}}^{(2)}= \frac{-igm_D}{\sqrt{2}m_W \cos{\beta}}
   U_{j2} \cos{\theta_U} e^{i\phi_D},
\eqno{(A.3.2)}
$$
$$
V_{D\tilde{U}_{2}\tilde{\chi}^{+}_{j}}^{(1)}= ig (V_{j1}^{\ast} \sin{\theta_U}
   e^{i\phi_U} - \frac{m_U}{\sqrt{2}m_W \sin{\beta}} V_{j2}^{\ast}
   \cos{\theta_U} e^{-i\phi_D}),
\eqno{(A.3.3)}
$$
$$
V_{D\tilde{U}_{2}\tilde{\chi}^{+}_{j}}^{(2)}= \frac{igm_D}{\sqrt{2}m_W \cos{\beta}}
   U_{i2} \sin{\theta_U} e^{-i\phi_D},
\eqno{(A.3.4)}
$$
  For the Feynman rules of the Higgs-quark-quark, Higgs-squark-squark,
Higgs-chargino-chargino and Z$(\gamma)$-chargino-chargino, one can refer
to Ref.\cite{haber}\cite{Gunion}.
The couplings of $Higgs(B)-\tilde{\chi}^{+}_{k}-\tilde{\chi}^{+}_{k}$ are
\begin{eqnarray*}
V_{B\tilde{\chi}^{+}_{k}\tilde{\chi}^{+}_{k}}=
V_{B\tilde{\chi}^{+}_{k}\tilde{\chi}^{+}_{k}}^{s}+
V_{B\tilde{\chi}^{+}_{k}\tilde{\chi}^{+}_{k}}^{ps} \gamma_5\hskip 5mm
(B=h^0, H^0, A^0, G^0), \hskip 10mm (A.4.1)
\end{eqnarray*}
where the notations defined above, which are involved in our calculation,
are explicitly expressed as below:
$$
V_{H^0\tilde{\chi}^{+}_{k}\tilde{\chi}^{+}_{k}}^{s} =
  \frac{-i g}{\sqrt{2}} \left[ \cos\alpha Re(V_{k,1} U_{k,2}) +
  \sin\alpha Re(V_{k,2} U_{k,1}) \right]
\eqno{(A.4.2)}
$$
$$
V_{h^0\tilde{\chi}^{+}_{k}\tilde{\chi}^{+}_{k}}^{s} =
   \frac{i g}{\sqrt{2}} \left[ \sin\alpha Re(V_{k,1} U_{k,2}) -
   \cos\alpha Re(V_{k,2} U_{k,1}) \right]
\eqno{(A.4.3)}
$$
$$
V_{A^0\tilde{\chi}^{+}_{k}\tilde{\chi}^{+}_{k}}^{ps} =
   \frac{g}{\sqrt{2}} \left[ \sin\beta Re(V_{k,1} U_{k,2}) +
   \cos\beta Re(V_{k,2} U_{k,1}) \right]
\eqno{(A.4.4)}
$$
$$
V_{G^0\tilde{\chi}^{+}_{k}\tilde{\chi}^{+}_{k}}^{ps} =
   \frac{-g}{\sqrt{2}} \left[ \cos\beta Re(V_{k,1} U_{k,2}) -
   \sin\beta Re(V_{k,2} U_{k,1}) \right]
\eqno{(A.4.5)}
$$
We define the following notations in Higgs-quark-quark couplings:
$$
H^0-U-U:\hskip 3mm
V_{H^{0}UU} = \frac{-i g m_U \sin{\alpha}}{2 m_W \sin{\beta}},\hskip 5mm
H^0-D-D:\hskip 3mm
V_{H^{0}DD} = \frac{-i g m_D \cos{\alpha}}{2 m_W \cos{\beta}},
\eqno{(A.5.1)}
$$
$$
h^0-U-U:\hskip 3mm
V_{h^{0}UU} = \frac{-i g m_U \cos{\alpha}}{2 m_W \sin{\beta}},\hskip 5mm
h^0-D-D:\hskip 3mm
V_{h^{0}DD} = \frac{i g m_D \sin{\alpha}}{2 m_W \cos{\delta}},
\eqno{(A.5.2)}
$$
$$
A^0-U-U:\hskip 3mm
V_{A^{0}UU}\gamma_5 = \frac{-g m_U \cot{\beta}}{2 m_W}\gamma_5,\hskip 5mm
A^0-D-D:\hskip 3mm
V_{A^{0}DD}\gamma_5 = \frac{-g m_D \tan{\beta}}{2 m_W}\gamma_5,
\eqno{(A.5.3)}
$$
$$
G^0-U-U:\hskip 3mm
V_{G^{0}UU}\gamma_5 = \frac{-g m_U}{2 m_W}\gamma_5,\hskip 5mm
G^0-D-D:\hskip 3mm
V_{G^{0}DD}\gamma_5 = \frac{g m_D}{2 m_W}\gamma_5.
\eqno{(A.5.4)}
$$
The couplings of $H^0(h^0)-\tilde{q}_{i}-\tilde{q}_{i}\hskip 3mm (i=1,2,q=u,d,c,s,t,b)$
are
\begin{eqnarray*}
V_{H^0\tilde{U}_{1}\tilde{U}_{1}} &=&
  \frac{-i g m_Z \cos{(\alpha+\beta)}}{\cos{\theta_W}} \left[
   (\frac{1}{2} - \frac{2}{3} \sin^2 \theta_W) \cos^2\theta_U +
    \frac{2}{3} \sin^2 \theta_W \sin^2\theta_U \right] \\
&& - \frac{i g m_U^2 \sin{\alpha}}{m_W \sin\beta}
   + \frac{i g m_U}{2 m_W \sin{\beta}} (A_U \sin{\alpha} + \mu \cos{\alpha})
     \sin{\theta_U} \cos{\theta_U} \cos{ 2 \phi_U},\hskip 5mm (A.6.1)
\end{eqnarray*}
\begin{eqnarray*}
V_{H^0\tilde{U}_{2}\tilde{U}_{2}} &=&
  \frac{-i g m_Z \cos(\alpha+\beta)}{\cos\theta_W} \left[
 (\frac{1}{2} - \frac{2}{3} \sin^2 \theta_W) \sin^2\theta_U +
  \frac{2}{3} \sin^2 \theta_W \cos^2\theta_U \right] \\
&& - \frac{i g m_U^2 \sin\alpha}{m_W \sin\beta}
   - \frac{i g m_U}{2 m_W \sin\beta} (A_U \sin\alpha + \mu \cos\alpha)
     \sin\theta_U \cos\theta_U \cos 2 \phi_U,\hskip 5mm (A.6.2)
\end{eqnarray*}
\begin{eqnarray*}
V_{H^0\tilde{D}_{1}\tilde{D}_{1}} &=&
  \frac{i g m_Z \cos(\alpha+\beta)}{\cos\theta_W} \left[
 (\frac{1}{2} - \frac{1}{3} \sin^2 \theta_W) \cos^2\theta_D +
  \frac{1}{3} \sin^2 \theta_W \sin^2\theta_D \right] \\
&& - \frac{i g m_D^2 \cos\alpha}{m_W \cos\beta}
   + \frac{i g m_D}{2 m_W \cos\beta} (A_D \cos\alpha + \mu \sin\alpha)
     \sin\theta_D \cos\theta_D \cos 2 \phi_D,\hskip 5mm (A.6.3)
\end{eqnarray*}
\begin{eqnarray*}
V_{H^0\tilde{D}_{2}\tilde{D}_{2}} &=&
  \frac{i g m_Z \cos(\alpha+\beta)}{\cos\theta_W} \left[
 (\frac{1}{2} - \frac{1}{3} \sin^2 \theta_W) \sin^2\theta_D +
  \frac{1}{3} \sin^2 \theta_W \cos^2\theta_D \right] \\
&& - \frac{i g m_D^2 \cos\alpha}{m_W \cos\beta}
   - \frac{i g m_D}{2 m_W \cos\beta} (A_D \cos\alpha + \mu \sin\alpha)
     \sin\theta_D \cos\theta_D \cos 2 \phi_D,\hskip 5mm (A.6.4)
\end{eqnarray*}
\begin{eqnarray*}
V_{h^0\tilde{U}_{1}\tilde{U}_{1}} &=&
  \frac{i g m_Z \sin(\alpha+\beta)}{\cos\theta_W} \left[
 (\frac{1}{2} - \frac{2}{3} \sin^2 \theta_W) \cos^2\theta_U +
  \frac{2}{3} \sin^2 \theta_W \sin^2\theta_U \right] \\
&& - \frac{i g m_U^2 \cos\alpha}{m_W \sin\beta}
   + \frac{i g m_U}{2 m_W \sin\beta} (A_U \cos\alpha - \mu \sin\alpha)
     \sin\theta_U \cos\theta_U \cos 2 \phi_U,\hskip 5mm (A.6.5)
\end{eqnarray*}
\begin{eqnarray*}
V_{h^0\tilde{U}_{2}\tilde{U}_{2}} &=&
  \frac{i g m_Z \sin(\alpha+\beta)}{\cos\theta_W} \left[
 (\frac{1}{2} - \frac{2}{3} \sin^2 \theta_W) \sin^2\theta_U +
  \frac{2}{3} \sin^2 \theta_W \cos^2\theta_U \right] \\
&& - \frac{i g m_U^2 \cos\alpha}{m_W \sin\beta}
   - \frac{i g m_U}{2 m_W \sin\beta} (A_U \cos\alpha - \mu \sin\alpha)
     \sin\theta_U \cos\theta_U \cos 2 \phi_U,\hskip 5mm (A.6.6)
\end{eqnarray*}
\begin{eqnarray*}
V_{h^0\tilde{D}_{1}\tilde{D}_{1}} &=&
 \frac{-i g m_Z \sin(\alpha+\beta)}{\cos\theta_W} \left[
(\frac{1}{2} - \frac{1}{3} \sin^2 \theta_W) \cos^2\theta_D +
 \frac{1}{3} \sin^2 \theta_W \sin^2\theta_D \right] \\
&& + \frac{i g m_D^2 \sin\alpha}{m_W \cos\beta}
   - \frac{i g m_D}{2 m_W \cos\beta} (A_D \sin\alpha - \mu \cos\alpha)
     \sin\theta_D \cos\theta_D \cos 2 \phi_D,\hskip 5mm (A.6.7)
\end{eqnarray*}
\begin{eqnarray*}
V_{h^0\tilde{D}_{2}\tilde{D}_{2}} &=&
  \frac{-i g m_Z \sin(\alpha+\beta)}{\cos\theta_W} \left[
 (\frac{1}{2} - \frac{1}{3} \sin^2 \theta_W) \sin^2\theta_D +
  \frac{1}{3} \sin^2 \theta_W \cos^2\theta_D \right] \\
&& + \frac{i g m_D^2 \sin\alpha}{m_W \cos\beta}
   + \frac{i g m_D}{2 m_W \cos\beta} (A_D \sin\alpha - \mu \cos\alpha)
     \sin\theta_D \cos\theta_D \cos 2 \phi_D,\hskip 5mm (A.6.8)
\end{eqnarray*}
respectively.
\par
B. Form Factors.
\par
As mentioned above, the amplitude parts from the u-channel box
and triangle vertex interaction diagrams can be obtained from the
t-channel's, so we present only the t-channel form factors for box and
triangle diagrams.
Since the form factors of the first and second generation (s)quarks
are analogous to those of the third generation (s)quarks,
here we list only the form factors of the box, triangle
and quartic interaction parts for the third generation quarks and squarks.
Actually we should take the sum of the form factors of each generation
(s)quarks for the total form factors.
In appendix, we use the notations defined in below for abbreviation.
$$
\bar{B}_{0}^{1,k}=B_{0}[-p_1-p_2, m_{\tilde{t}_k},m_{\tilde{t}_k}]-\Delta, \hskip 8mm
\bar{B}_{0}^{2,k}=B_{0}[-p_1-p_2, m_{\tilde{b}_k},m_{\tilde{b}_k}]-\Delta,
\eqno{(B.1)}
$$
$$
C_{0}^{1},C_{ij}^{1}=
    C_{0},C_{ij}[p_3,-p_1-p_2,m_t,m_t,m_t],
$$
$$
C_{0}^{2},C_{ij}^{2}=
    C_{0},C_{ij}[p_3,-p_1-p_2,m_b,m_b,m_b],
$$
$$
C_{0}^{3,k},C_{ij}^{3,k}=
    C_{0},C_{ij}[p_3,-p_1-p_2,m_{\tilde{t}_k},m_{\tilde{t}_k},m_{\tilde{t}_k}],
$$
$$
C_{0}^{4,k},C_{ij}^{4,k}=
    C_{0},C_{ij}[p_3,-p_1-p_2,m_{\tilde{b}_k},m_{\tilde{b}_k},m_{\tilde{b}_k}],
$$
$$
C_{0}^{5,k},C_{ij}^{5,k}=
    C_{0},C_{ij}[-p_1,p_1+p_2,m_t,m_{\tilde{b}_k},m_{\tilde{b}_k}],
$$
$$
C_{0}^{6,k},C_{ij}^{6,k}=
    C_{0},C_{ij}[-p_1,p_1+p_2,m_b,m_{\tilde{t}_k},m_{\tilde{t}_k}],
$$
$$
D_{0}^{1,k},D_{ij}^{1,k},D_{ijl}^{1,k}=
    D_{0},D_{ij},D_{ijl}[-p_1,p_3,p_4,m_{\tilde{b}_k},m_t,m_t,m_t]
$$
$$
D_{0}^{2,k},D_{ij}^{2,k},D_{ijl}^{2,k}=
    D_{0},D_{ij},D_{ijl}[-p_1,p_3,p_4,m_{\tilde{t}_k},m_b,m_b,m_b]
$$
$$
D_{0}^{3,k},D_{ij}^{3,k},D_{ijl}^{3,k}=
    D_{0},D_{ij},D_{ijl}[-p_1,p_3,p_4,
                m_t,m_{\tilde{b}_k},m_{\tilde{b}_k},m_{\tilde{b}_k}]
$$
$$
D_{0}^{4,k},D_{ij}^{4,k},D_{ijl}^{4,k}=
   D_{0},D_{ij},D_{ijl}[-p_1,p_3,p_4,
                m_b,m_{\tilde{t}_k},m_{\tilde{t}_k},m_{\tilde{t}_k}]
$$
$$
D_{0}^{5,k},D_{ij}^{5,k},D_{ijl}^{5,k}=
    D_{0},D_{ij},D_{ijl}[p_3,p_2-p_3,-p_4,m_t,m_t,m_{\tilde{b}_k},m_{\tilde{b}_k}]
$$
$$
D_{0}^{6,k},D_{ij}^{6,k},D_{ijl}^{6,k}=
    D_{0},D_{ij},D_{ijl}[p_3,p_2-p_3,-p_4,m_b,m_b,m_{\tilde{t}_k},m_{\tilde{t}_k}]
\eqno{(B.2)}
$$
\begin{eqnarray*}
A_{t}=\frac{i}{\hat{t}-m_{\tilde{\chi}^{+}_1}^2},\hskip 5mm
A_{u}=\frac{i}{\hat{u}-m_{\tilde{\chi}^{+}_1}^2},
\end{eqnarray*}
\begin{eqnarray*}
A_{h}=\frac{i}{\hat{s}-m_{h}^2},\hskip 5mm
A_{H}=\frac{i}{\hat{s}-m_{H}^2}.
\end{eqnarray*}
\begin{eqnarray*}
A_{A}=\frac{i}{\hat{s}-m_{A}^2},\hskip 5mm
A_{G}=\frac{i}{\hat{s}-m_{Z}^2}. \hskip 5mm (B.3)
\end{eqnarray*}
$$
F_{1,k}=
    -V_{t\tilde{b}_{k}\tilde{\chi}^{+}_{1}}^{(1)}
    V_{t\tilde{b}_{k}\tilde{\chi}^{+}_{1}}^{(2)\ast} -
    V_{t\tilde{b}_{k}\tilde{\chi}^{+}_{1}}^{(2)}
    V_{t\tilde{b}_{k}\tilde{\chi}^{+}_{1}}^{(1)\ast},\hskip 5mm
F_{2,k}=
   -|V_{t\tilde{b}_{k}\tilde{\chi}^{+}_{1}}^{(1)}|^2-
   |V_{t\tilde{b}_{k}\tilde{\chi}^{+}_{1}}^{(2)}|^2
$$
$$
F_{3,k}=
    -V_{t\tilde{b}_{k}\tilde{\chi}^{+}_{1}}^{(1)}
    V_{t\tilde{b}_{k}\tilde{\chi}^{+}_{1}}^{(2)\ast} +
    V_{t\tilde{b}_{k}\tilde{\chi}^{+}_{1}}^{(2)}
    V_{t\tilde{b}_{k}\tilde{\chi}^{+}_{1}}^{(1)\ast},\hskip 5mm
F_{4,k}=
   -|V_{t\tilde{b}_{k}\tilde{\chi}^{+}_{1}}^{(1)}|^2+
   |V_{t\tilde{b}_{k}\tilde{\chi}^{+}_{1}}^{(2)}|^2
$$
$$
F_{5,k}=
    V_{b\tilde{t}_{k}\tilde{\chi}^{+}_{1}}^{(1)}
    V_{b\tilde{t}_{k}\tilde{\chi}^{+}_{1}}^{(2)\ast} +
    V_{b\tilde{t}_{k}\tilde{\chi}^{+}_{1}}^{(2)}
    V_{b\tilde{t}_{k}\tilde{\chi}^{+}_{1}}^{(1)\ast},\hskip 5mm
F_{6,k}=
   |V_{b\tilde{t}_{k}\tilde{\chi}^{+}_{1}}^{(1)}|^2 +
   |V_{b\tilde{t}_{k}\tilde{\chi}^{+}_{1}}^{(2)}|^2
$$
$$
F_{7,k}=
    V_{b\tilde{t}_{k}\tilde{\chi}^{+}_{1}}^{(1)}
    V_{b\tilde{t}_{k}\tilde{\chi}^{+}_{1}}^{(2)\ast} -
    V_{b\tilde{t}_{k}\tilde{\chi}^{+}_{1}}^{(2)}
    V_{b\tilde{t}_{k}\tilde{\chi}^{+}_{1}}^{(1)\ast}, \hskip 5mm
F_{8,k}=
   |V_{b\tilde{t}_{k}\tilde{\chi}^{+}_{1}}^{(1)}|^2 -
   |V_{b\tilde{t}_{k}\tilde{\chi}^{+}_{1}}^{(2)}|^2
\eqno{(B.4)}
$$
In the following, the expression denoted as $(t \rightarrow b)$ means
doing the replacements of $Q_t \rightarrow Q_b$, $m_t \rightarrow m_b$,
$F_{1,k} \rightarrow F_{5,k}$, $F_{2,k} \rightarrow F_{6,k}$,
$F_{3,k} \rightarrow F_{7,k}$, $F_{4,k} \rightarrow F_{8,k}$,
$B_{0}^{1,k} \rightarrow B_{0}^{2,k}$,
$C^{1} \rightarrow C^{2}$, $C^{3,k} \rightarrow C^{4,k}$,
$C^{5,k} \rightarrow C^{6,k}$, $D^{1,k} \rightarrow D^{2,k}$,
$D^{3,k} \rightarrow D^{4,k}$, $D^{5,k} \rightarrow D^{6,k}$.
The form factors in the amplitude of the quartic interaction
diagrams Fig.1(b) are expressed as:
\begin{eqnarray*}
f_{1}^{q}= f_{2}^{q} &=&
\frac{g_s^2}{32 \pi^2} \sum_{k=1,2} \left[ 2 \bar{B}_0^{1,k} (A_h V_{h^0\tilde{t}_{k}
     \tilde{t}_{k}} V_{h^0\tilde{\chi}^{+}_{1}\tilde{\chi}^{+}_{1}}^{s}
     +  A_H V_{H^0\tilde{t}_{k}\tilde{t}_{k}} V_{H^0\tilde{\chi}^{+}_{1}
     \tilde{\chi}^{+}_{1}}^{s}) \right. \\
&-&  \left. i (C_0^{5,k} m_t F_{1,k} - C_{11}^{5,k} m_{\tilde{\chi}_{1}^{+}}
     F_{2,k}) \right ] + (t \rightarrow b)
\end{eqnarray*}
\begin{eqnarray*}
f_{34}^{q} = f_{35}^{q} &=&
\frac{g_s^2}{32 \pi^2} \sum_{k=1,2} \left\{ 2\bar{B}_0^{1,k} (A_h V_{h^0\tilde{t}_{k}
     \tilde{t}_{k}} V_{h^0\tilde{\chi}^{+}_{1}\tilde{\chi}^{+}_{1}}^{ps}
     + A_H V_{H^0\tilde{t}_{k}\tilde{t}_{k}} V_{H^0\tilde{\chi}^{+}_{1}
     \tilde{\chi}^{+}_{1}}^{ps}) \right. \\
&+&  \left. i [C_0^{5,k} m_t F_{3,k} +(2 C_{12}^{5,k} - C_{11}^{5,k})
     m_{\tilde{\chi}_{1}^{+}} F_{4,k}] \right\} + (t \rightarrow b)
\end{eqnarray*}
\begin{eqnarray*}
f_{i}^{q} = 0.\hskip 10mm  (i=3 \sim 33, 36 \sim 66)
\end{eqnarray*}
\par
The form factors in the amplitude from the t-channel triangle
diagrams depicted in Fig.1(c), are listed below:
\begin{eqnarray*}
f_{1}^{tr,\hat{t}} &=& f_{2}^{tr,\hat{t}} \\
&=& \frac{g_s^2}{8 \pi^2} \left\{m_t (A_h V_{h^0tt} V_{h^0\tilde{\chi}_1^{+}
    \tilde{\chi}_1^{+}}^s + A_H V_{H^0tt} V_{H^0\tilde{\chi}_1^{+}\tilde
    {\chi}_1^{+}}^s ) \left[- C_0^1 m_t^2 + 2 C_{22}^1
    (p_1 \cdot p_2  + m_{\tilde{\chi}_1^{+}}^2) \right.\right.\\
&+& \left. (C_0^1 - 2 C_{23}^1) (p_1 + p_2) \cdot p_3 \right] -
    \sum_{k=1,2} 2\bar{C}_{24}^{3,k} (A_h V_{h^0 \tilde{t}_k\tilde{t}_k}
    V_{h^0\tilde{\chi}_1^{+}\tilde{\chi}_1^{+}}^s + \left. A_H V_{H^0\tilde{t}
    _k\tilde{t}_k} V_{H^0\tilde{\chi}_1^{+}\tilde{\chi}_ 1^{+}}^s) \right\} \\
&+& (t \rightarrow b)
\end{eqnarray*}
\begin{eqnarray*}
f_3^{tr,\hat{t}} &=& f_4^{tr,\hat{t}} =
\frac{g_s^2}{4 \pi^2} (\frac{i e^2 Q_t}{\hat{s}} - A_G V_{Z^0tt}^v
    V_{Z^0 \tilde{\chi}_1^{+}\tilde{\chi}_1^{+}}^s) [2 (\bar{C}_{24}^1 +
    \bar{C}_{36}^1 - \bar{C}_{35}^1) \\
&+& (C_0^1 + C_{12}^1- C_{11}^1 ) m_t^2 + 2 (C_{34}^1 - C_{12}^1 - C_{22}^1
    - C_{32}^1) (p_1 \cdot p_2 + m_{\tilde{\chi}_1^{+}}^2) \\
&+& 2 (C_{12}^1 + C_{22}^1 + C_{34}^1- C_{33}^1 ) (p_1 + p_2 )\cdot p_3]
    +(t \rightarrow b)
\end{eqnarray*}
\begin{eqnarray*}
f_5^{tr,\hat{t}} &=& f_6^{tr,\hat{t}} =
-\frac{g_s^2}{4 \pi^2} (\frac{i e^2 Q_t}{\hat{s}} - A_G V_{Z^0tt}^v
    V_{Z^0 \tilde{\chi}_1^{+}\tilde{\chi}_1^{+}}^s) [2 \bar{C}_{24}^1 -
    2 \bar{C}_{36}^1 + (C_0^1 - C_{12}^1) m_t^2 \\
&+& 2 (C_{22}^1 + C_{32}^1) (p_1 \cdot p_2 + m_{\tilde{\chi}_1^{+}}^2)
   - 2 (C_{22}^1 + C_{34}^1) (p_1 + p_2) \cdot p_3] +(t \rightarrow b)
\end{eqnarray*}
\begin{eqnarray*}
f_7^{tr,\hat{t}} &=& f_8^{tr,\hat{t}} = f_9^{tr,\hat{t}} = f_{10}^{tr,\hat{t}}=
\frac{g_s^2} {4 \pi^2} \left[ m_t (A_h V_{h^0tt} V_{h^0\tilde{\chi}_1^{+}
     \tilde{\chi}_1^{+}}^s + A_H V_{H^0tt} V_{H^0\tilde{\chi}_1^{+}\tilde
     {\chi}_1^{+}}^s) (4 C_{23}^1 - C_0^1 - 4 C_{22}^1) \right. \\
&+&  \sum_{k=1,2} \left. 2 (A_h V_{h^0\tilde{t}_k\tilde{t}_k} V_{h^0\tilde{\chi}_1^{+}
     \tilde{\chi}_1^{+}}^s + A_H V_{H^0\tilde{t}_kt\tilde{t}_k} V_{H^0\tilde
     {\chi}_1^{+}\tilde{\chi}_1^{+}}^s) (C_{22}^{3,k} - C_{23}^{3,k})
     \right] + (t \rightarrow b)
\end{eqnarray*}
\begin{eqnarray*}
f_{11}^{tr,\hat{t}} &=& f_{12}^{tr,\hat{t}} =
\frac{g_s^2}{8 \pi^2} (\frac{i e^2 Q_t}{\hat{s}} - A_G V_{Z^0tt}^v
    V_{Z^0 \tilde{\chi}_1^{+}\tilde{\chi}_1^{+}}^s)
    [-4 \bar{C}_{24}^1 - 2 \bar{C}_{35}^1  \\
&-& (2 C_0^1 + C_{11}^1) m_t^2 + 2 (C_{12}^1 + C_{34}^1 + 2 C_{22}^1 )
    (p_1 \cdot p_2 + m_{\tilde{\chi}_1^+}^2) \\
&-& 2 (C_{12}^1  + C_{33}^1 + 2 C_{23}^1) (p_1 + p_2 )\cdot p_3]
    + (t \rightarrow b)
\end{eqnarray*}
\begin{eqnarray*}
f_{17}^{tr,\hat{t}} = f_{18}^{tr,\hat{t}} = f_{19}^{tr,\hat{t}}
= f_{20}^{tr,\hat{t}} &=&
-\frac{g_s^2}{\pi^2} (\frac{i e^2 Q_t}{4 \hat{s}} - A_G V_{Z^0tt}^v
    V_{Z^0 \tilde{\chi}_1^{+}\tilde{\chi}_1^{+}}^s)
    (C_{22}^1 + C_{34}^1 - C_{23}^1 -C_{33}^1) \\
&+& (t \rightarrow b)
\end{eqnarray*}
\begin{eqnarray*}
f_{21}^{tr,\hat{t}} = f_{22}^{tr,\hat{t}} =
-\frac{i g_s^2 m_t}{4 \pi^2} C_0^1
    (A_A V_{A^0tt} V_{A^0\tilde{\chi}_1^{+}\tilde{\chi}_1^{+}}^s +
    A_G V_{G^0tt} V_{G^0\tilde{\chi}_1^{+}\tilde{\chi}_1^{+}}^s)
    +(t \rightarrow b)
\end{eqnarray*}
\begin{eqnarray*}
f_{23}^{tr,\hat{t}} &=& f_{24}^{tr,\hat{t}} =
\frac{i g_s^2}{4 \pi^2} A_G V_{Z^0tt}^{pv} V_{Z^0 \tilde{\chi}_1^{+}
    \tilde{\chi}_1^{+}}^s [-2 \bar{C}_{24}^1 - 6 \bar{C}_{36}^1 -
    (C_0^1 + C_{12}^1) m_t^2 \\
&+& 2 (C_{22}^1 + C_{32}^1) (p_1 \cdot p_2 + m_{\tilde{\chi}_1^{+}}^2)
    - 2 (C_{23}^1 + C_{34}^1) (p_1 + p_2 )\cdot p_3] +(t \rightarrow b)
\end{eqnarray*}
\begin{eqnarray*}
f_{25}^{tr,\hat{t}} &=&
-\frac{i g_s^2}{4 \pi^2} A_G V_{Z^0tt}^{pv} V_{Z^0 \tilde{\chi}_1^{+}
    \tilde{\chi}_1^{+}}^s [-4 \bar{C}_{24}^1 - 6 \bar{C}_{35}^1 -
    (2 C_0^1 + C_{11}^1) m_t^2 \\
&+& 2 (C_{12}^1 + C_{34}^1 + 2 C_{22}^1) (p_1 \cdot p_2 + m_{\tilde{\chi}
    _1^{+}}^2) - 2 (C_{12}^1 + C_{33}^1 + 2 C_{23}^1) (p_1 + p_2 )\cdot p_3] \\
&+& (t \rightarrow b)
\end{eqnarray*}
\begin{eqnarray*}
f_{26}^{tr,\hat{t}} = f_{27}^{tr,\hat{t}} = f_{28}^{tr,\hat{t}}
= f_{29}^{tr,\hat{t}} =
\frac{i g_s^2}{2 \pi^2} A_G V_{Z^0tt}^{pv} V_{Z^0 \tilde{\chi}_1^{+}\tilde
    {\chi} _1^{+}}^s (C_{22}^1 - C_{23}^1) +(t \rightarrow b)
\end{eqnarray*}
\begin{eqnarray*}
f_{30}^{tr,\hat{t}} = f_{31}^{tr,\hat{t}} = f_{32}^{tr,\hat{t}}
= f_{33}^{tr,\hat{t}} =-f_{26}^{tr,\hat{t}}
\end{eqnarray*}
\begin{eqnarray*}
f_{34}^{tr,\hat{t}} &=& f_{35}^{tr,\hat{t}} =
-\frac{g_s^2}{4 \pi^2} \{ \sum_{k=1,2} \bar{C}_{24}^{3,k} (A_h V_{h^0\tilde{t}
    _k\tilde{t}_k} V_{h^0\tilde{\chi}_1^{+}\tilde{\chi}_1^{+}}^{ps} +
    A_H V_{H^0\tilde{t}_kt\tilde{t}_k} V_{H^0\tilde{\chi}_1^{+}\tilde{\chi}_
    1^{+}}^{ps}) \\
&+& A_G V_{Z^0tt}^v V_{Z^0 \tilde{\chi}_1^{+}\tilde{\chi}_1^{+}}^{ps}
    [2 \bar{C}_{24}^1 + 2 \bar{C}_{36}^1 +(C_0^1 + C_{12}^1) m_t^2 \\
&-& 2 (C_{22}^1 + C_{32}^1) (p_1 \cdot p_2 + m_{\tilde{\chi}_1^{+}}^2)
    + 2 (C_{23}^1 +  C_{34}^1) (p_1 + p_2 )\cdot p_3] \}
    +(t \rightarrow b)
\end{eqnarray*}
\begin{eqnarray*}
f_{36}^{tr,\hat{t}} &=& f_{37}^{tr,\hat{t}} =
\frac{g_s^2}{4 \pi^2} A_G V_{Z^0tt}^v V_{Z^0 \tilde{\chi}_1^{+}\tilde{\chi}_1
    ^{+}}^{ps} [2 \bar{C}_{24}^1 + 2 \bar{C}_{36}^1 - 2 \bar{C}_{35}^1  \\
&+& (C_0^1 + C_{12}^1 - C_{11}^1 ) m_t^2 + 2 (C_{34}^1 - C_{12}^1 - C_{22}^1
    - C_{32}^1 ) (p_1 \cdot p_2 + m_{\tilde{\chi}_1^{+}}^2) \\
&+& 2 (C_{12}^1 + C_{22}^1 + C_{34}^1 - C_{33}^1 ) (p_1 + p_2 )\cdot p_3]
    +(t \rightarrow b)
\end{eqnarray*}
\begin{eqnarray*}
f_{38}^{tr,\hat{t}} &=& f_{39}^{tr,\hat{t}} =
-\frac{g_s^2}{4 \pi^2} A_G V_{Z^0tt}^v V_{Z^0 \tilde{\chi}_1^{+}\tilde{\chi}
    _1^{+}}^{ps} [2 \bar{C}_{24}^1 - 2 \bar{C}_{36}^1 +
    (C_0^1 - C_{12}^1) m_t^2 \\
&+& 2 (C_{22}^1 + C_{32}^1) (p_1 \cdot p_2 + m_{\tilde{\chi}_1^{+}}^2)
    - 2 (C_{22}^1 + C_{34}^1) (p_1 + p_2 )\cdot p_3] +(t \rightarrow b)
\end{eqnarray*}
\begin{eqnarray*}
f_{40}^{tr,\hat{t}} &=& f_{41}^{tr,\hat{t}} = f_{42}^{tr,\hat{t}}
         = f_{43}^{tr,\hat{t}}=
\frac{g_s^2}{4 \pi^2} \left\{ m_t
    (A_h V_{h^0tt} V_{h^0\tilde{\chi}_1^{+}\tilde{\chi}_1^{+}}^{ps} +
    A_H V_{H^0tt} V_{H^0\tilde{\chi}_1^{+}\tilde{\chi}_1^{+}}^{ps})
    (4 C_{23}^1 - C_{0}^1 - 4 C_{22}^1) \right. \\
&+& 4 A_G V_{Z^0tt}^v V_{Z^0 \tilde{\chi}_1^{+}\tilde
    {\chi}_1^{+}}^{ps} (C_{23}^1 + 2 C_{34}^1 - C_{22}^1 - 2 C_{32}^1) \\
&+& \sum_{k=1,2} \left. 2 (C_{22}^{3,k} - C_{23}^{3,k})
    (A_h V_{h^0\tilde{t}_k\tilde{t}_k} V_{h^0\tilde{\chi}_1^{+}\tilde{\chi}_1^{+}}^{ps} +
    A_H V_{H^0\tilde{t}_kt\tilde{t}_k} V_{H^0\tilde{\chi}_1^{+}\tilde{\chi}_1^{+}}^{ps})
    \right\} + (t \rightarrow b)
\end{eqnarray*}
\begin{eqnarray*}
f_{44}^{tr,\hat{t}} &=& f_{45}^{tr,\hat{t}} =
\frac{g_s^2}{8 \pi^2} A_G V_{Z^0tt}^v V_{Z^0 \tilde{\chi}_1^{+}\tilde{\chi}
    _1^{+}}^{ps} [-4 \bar{C}_{24}^1 - 2 \bar{C}_{35}^1 -
    (2 C_0^1 + C_{11}^1) m_t^2 \\
&+& 2 (C_{12}^1 + C_{34}^1+ 2 C_{22}^1) (p_1 \cdot p_2 + m_{\tilde{\chi}_1^+}^2)
    - 2 (C_{12}^1 + C_{33}^1+ 2 C_{23}^1) (p_1 + p_2 )\cdot p_3] \\
&+& (t \rightarrow b)
\end{eqnarray*}
\begin{eqnarray*}
f_{50}^{tr,\hat{t}} = f_{51}^{tr,\hat{t}} = f_{52}^{tr,\hat{t}}
= f_{53}^{tr,\hat{t}} =
-\frac{g_s^2}{\pi^2} A_G V_{Z^0tt}^v V_{Z^0 \tilde{\chi}_1^{+}\tilde{\chi}_1^
    {+}}^{ps} (C_{22}^1 + C_{34}^1 - C_{23}^1 -C_{33}^1 ) +(t \rightarrow b)
\end{eqnarray*}
\begin{eqnarray*}
f_{54}^{tr,\hat{t}} = f_{55}^{tr,\hat{t}} =
-\frac{i g_s^2 m_t}{4 \pi^2} C_{0}^1
    (A_A V_{A^0tt} V_{A^0\tilde{\chi}_1^{+}\tilde{\chi}_1^{+}}^{ps} +
    A_G V_{G^0tt} V_{G^0\tilde{\chi}_1^{+}\tilde{\chi}_1^{+}}^{ps})
    +(t \rightarrow b)
\end{eqnarray*}
\begin{eqnarray*}
f_{56}^{tr,\hat{t}} &=& f_{57}^{tr,\hat{t}} =
-\frac{i g_s^2}{4 \pi^2} A_G V_{Z^0tt}^{pv} V_{Z^0 \tilde{\chi}_1^{+}\tilde
    {\chi}_1^{+}}^{ps} [-2 \bar{C}_{24}^1 - 6 \bar{C}_{36}^1 -
    (C_0^1 + C_{12}^1) m_t^2 \\
&+& 2 (C_{22}^1 + C_{32}^1) (p_1 \cdot p_2 + m_{\tilde{\chi}_1^{+}}^2)
    - 2 (C_{23}^1 + C_{34}^1) (p_1 + p_2 )\cdot p_3] +(t \rightarrow b)
\end{eqnarray*}
\begin{eqnarray*}
f_{58}^{tr,\hat{t}} &=&
\frac{i g_s^2}{4 \pi^2} A_G V_{Z^0tt}^{pv} V_{Z^0 \tilde{\chi}_1^{+}\tilde
    {\chi}_1^{+}}^{ps} [-4 \bar{C}_{24}^1 - 6 \bar{C}_{35}^1 -
    (2 C_0^1 + C_{11}^1) m_t^2 \\
&+& 2 (C_{12}^1 + 2 C_{22}^1 + C_{34}^1) (p_1 \cdot p_2 + m_{\tilde{\chi}_1
    ^{+}}^2) - 2 (C_{12}^1 + 2 C_{23}^1 + C_{33}^1) (p_1 + p_2 )\cdot p_3] \\
&+& (t \rightarrow b)
\end{eqnarray*}
\begin{eqnarray*}
f_{59}^{tr,\hat{t}} = f_{60}^{tr,\hat{t}} = f_{61}^{tr,\hat{t}}
= f_{62}^{tr,\hat{t}} =
\frac{i g_s^2}{2 \pi^2} A_G V_{Z^0tt}^{pv} V_{Z^0 \tilde{\chi}_1^{+}\tilde
    {\chi}_1^{+}}^{ps} (C_{23}^1 - C_{22}^1) +(t \rightarrow b)
\end{eqnarray*}
\begin{eqnarray*}
f_{63}^{tr,\hat{t}} = f_{64}^{tr,\hat{t}} = f_{65}^{tr,\hat{t}}
= f_{66}^{tr,\hat{t}} =-f_{59}^{tr,\hat{t}}
\end{eqnarray*}
\begin{eqnarray*}
f_{i}^{tr,\hat{t}} = 0 \hskip 10mm (i=13 \sim 16, 46 \sim 49)
\end{eqnarray*}
where $\bar{C}_{24}=C_{24}-\frac{\Delta}{4}$, $\bar{C}_{35}=
C_{35}+\frac{\Delta}{6}$ and $\bar{C}_{36}=C_{36}+\frac{\Delta}{12}$.
The form factors of the amplitude part from t-channel box diagrams
Fig.1(a) are written as:
\begin{eqnarray*}
f_1^{b,\hat{t}} &=&
\frac{i g_s^2}{32 \pi^2} \sum_{k=1,2} \left\{ m_t F_{1,k} \left[ 2(D_{23}^{1,k}
    -D_{13}^{1,k}-D_{25}^{1,k}) p_1 \cdot p_2+2(D_{25}^{1,k}+D_{26}^{1,k}
     \right. \right. +
    2D_{13}^{1,k}-D_{11}^{1,k} \\
&-& D_{12}^{1,k}-D_{23}^{1,k}-D_{24}^{1,k})
    p_1 \cdot p_3+2(D_{13}^{1,k}+D_{26}^{1,k}-D_{23}^{1,k})
    p_2 \cdot p_3+(2D_{11}^{1,k}+2 D_{23}^{1,k} \\
&+& D_{0}^{1,k}+D_{21}^{1,k}-2
    D_{13}^{1,k}-2D_{25}^{1,k})
    \left. m_{\tilde{\chi}_1^+}^2-4D_{27}^{1,k}+2(D_{27}^{5,k}+D_{27}^{1,k}+
    D_{27}^{3,k})-D_{0}^{1,k} m_t^2 \right]\\
&+& m_{\tilde{\chi}_1^+} F_{2,k} \left[ 2(D_{23}^{1,k}+D_{37}^{1,k}-
    D_{13}^{1,k}-D_{35}^{1,k}-2 D_{25}^{1,k})  \right.
    p_1 \cdot p_2+2(3D_{25}^{1,k}+D_{26}^{1,k}
    +D_{310}^{1,k} \\
&+& D_{35}^{1,k}+
    2D_{13}^{1,k}-D_{11}^{1,k}-D_{12}^{1,k}-D_{21}^{1,k}
    -D_{23}^{1,k}-D_{34}^{1,k}-D_{37}^{1,k}-2D_{24}^{1,k}) p_1 \cdot p_3 \\
&+& 2(D_{13}^{1,k}+D_{25}^{1,k}+D_{26}^{1,k}+D_{310}^{1,k}
    -D_{23}^{1,k}-D_{37}^{1,k}) p_2 \cdot p_3-(D_{0}^{1,k}+D_{11}^{1,k}) m_t^2\\
&-& 4(D_{27}^{1,k}
    + D_{311}^{1,k})-2(D_{312}^{5,k}+D_{311}^{3,k})\\
&+& (3D_{11}^{1,k}+3D_{21}^{1,k}+
    2D_{23}^{1,k}+2D_{37}^{1,k}+D_{0}^{1,k}+D_{31}^{1,k}
    - \left. \left. 2D_{13}^{1,k}-2D_{35}^{1,k}-4D_{25}^{1,k})
    m_{\tilde{\chi}_1^+}^2 \right] \right\}\\
&+& (t \rightarrow b)
\end{eqnarray*}
\begin{eqnarray*}
f_2^{b,\hat{t}} &=&
\frac{i g_s^2}{16 \pi^2} \sum_{k=1,2} \left[ m_t F_{1,k} (D_{27}^{1,k}+
    D_{27}^{3,k}+D_{27}^{5,k})+m_{\tilde{\chi}_1^+} F_{2,k} (D_{27}^{1,k}
     \right.
    + \left. D_{311}^{1,k}-D_{311}^{3,k}-D_{312}^{5,k}) \right]\\
&+& (t \rightarrow b)
\end{eqnarray*}
\begin{eqnarray*}
f_3^{b,\hat{t}} &=&
\frac{i g_s^2}{16 \pi^2} \sum_{k=1,2} F_{2,k} \left\{ 2(D_{26}^{1,k}+
    D_{310}^{1,k}+D_{37}^{1,k}-D_{25}^{1,k}-D_{35}^{1,k}-D_{39}^{1,k}+
    D_{37}^{5,k} \right.
    +D_{38}^{5,k}
    -D_{310}^{5,k} \\
&-& D_{39}^{5,k}) p_1 \cdot p_2+2(D_{22}^{1,k}+
    D_{25}^{1,k}+D_{35}^{1,k}+D_{36}^{1,k}+D_{39}^{1,k}
    - D_{24}^{1,k}-D_{26}^{1,k}
    -D_{34}^{1,k}-D_{37}^{1,k} \\
&-& D_{38}^{1,k}+2
    D_{310}^{5,k}+D_{26}^{5,k}+D_{39}^{5,k}-D_{25}^{5,k}
    -D_{35}^{5,k}-D_{37}^{5,k}-D_{38}^{5,k}) p_1 \cdot p_3
    +2(D_{25}^{1,k}+
    D_{310}^{1,k} \\
&+& D_{39}^{1,k}-D_{26}^{1,k}-D_{37}^{1,k}
    -D_{38}^{1,k}+3D_{310}^{5,k}-2D_{36}^{5,k}-2D_{38}^{5,k}+D_{32}^{5,k}+
    D_{34}^{5,k}+D_{39}^{5,k}-D_{35}^{5,k}\\
&-& D_{37}^{5,k})
    p_2 \cdot p_3+(D_{12}^{1,k}-D_{11}^{1,k}+D_{12}^{5,k}-D_{11}^{5,k}) m_t^2+
    (D_{11}^{1,k}+D_{31}^{1,k}+2D_{21}^{1,k}
    +2D_{26}^{1,k}\\
&+& 2D_{310}^{1,k}+2D_{37}^{1,k}-D_{12}^{1,k}-D_{34}^{1,k}-2
    D_{24}^{1,k}-2D_{25}^{1,k}-2D_{35}^{1,k}-2 D_{39}^{1,k}
    + D_{36}^{5,k}-D_{32}^{5,k}\\
&+& 2 D_{37}^{5,k}+2 D_{38}^{5,k}-2 D_{310}^{5,k}-2
    D_{39}^{5,k}) m_{\tilde{\chi}_1^+}^2 \\
&+& 4(D_{312}^{1,k}-D_{311}^{1,k}+ D_{312}^{5,k}-D_{311}^{5,k})
    + \left. 2(D_{312}^{3,k}-D_{311}^{3,k}) \right\}
    +(t \rightarrow b)
\end{eqnarray*}
\begin{eqnarray*}
f_4^{b,\hat{t}} &=&
\frac{i g_s^2}{16 \pi^2} \sum_{k=1,2}
    F_{2,k} \left\{ 2(D_{26}^{1,k}+D_{310}^{1,k}+D_{33}^{1,k}-
    D_{23}^{1,k}-D_{37}^{1,k} \right.
    -D_{39}^{1,k}+D_{23}^{5,k}+D_{37}^{5,k}-D_{26}^{5,k}\\
&-& D_{310}^{5,k})
    p_1 \cdot p_2+2\left[ 2(D_{23}^{1,k}+D_{39}^{1,k}-D_{26}^{1,k} \right.
    -D_{310}^{1,k})+D_{22}^{1,k}+D_{36}^{1,k}+D_{37}^{1,k}-D_{25}^{1,k}
    -D_{33}^{1,k}\\
&-& D_{38}^{1,k}+D_{26}^{5,k}+D_{310}^{5,k}
    - \left. D_{13}^{5,k}-D_{23}^{5,k}-D_{35}^{5,k}-D_{37}^{5,k}-2D_{25}^{5,k}
     \right] p_1 \cdot p_3+2(2D_{39}^{1,k}+D_{23}^{1,k}\\
&-& D_{26}^{1,k}-
    D_{33}^{1,k}-D_{38}^{1,k}+2D_{26}^{5,k}+2D_{310}^{5,k}+D_{24}^{5,k}+
    D_{34}^{5,k}-D_{22}^{5,k}-D_{23}^{5,k}-D_{25}^{5,k}
    -D_{35}^{5,k}\\
&-& D_{36}^{5,k}-D_{37}^{5,k}) p_2 \cdot p_3+(D_{12}^{1,k}-
    D_{13}^{1,k}-D_{0}^{5,k}-D_{11}^{5,k})
    m_t^2+\left[ 2(D_{25}^{1,k}+D_{26}^{1,k}+D_{310}^{1,k} \right.\\
&+& D_{33}^{1,k}-
    D_{23}^{1,k}-D_{24}^{1,k}-D_{37}^{1,k}-D_{39}^{1,k}
    +D_{23}^{5,k}+D_{37}^{5,k}-D_{26}^{5,k}-D_{310}^{5,k})+D_{13}^{1,k}+
    D_{35}^{1,k}\\
&-& D_{12}^{1,k}-D_{34}^{1,k}+D_{22}^{5,k}+
    \left. D_{36}^{5,k} \right] m_{\tilde{\chi}_1^+}^2 +
    4D_{312}^{1,k}+2D_{27}^{1,k}-6 D_{313}^{1,k}\\
&+& \left. 2D_{27}^{3,k}+2D_{312}^{3,k}-2D_{27}^{5,k}-4D_{311}^{5,k} \right\}
    +(t \rightarrow b)
\end{eqnarray*}
\begin{eqnarray*}
f_5^{b,\hat{t}} &=&
\frac{i g_s^2}{16 \pi^2} \sum_{k=1,2} \left\{ 2m_t m_{\tilde{\chi}_1^+} F_{1,k}
     (D_{13}^{1,k}-D_{0}^{1,k}-D_{11}^{1,k}) \right. +
    F_{2,k} \left[ 2(D_{26}^{1,k}-D_{25}^{1,k}) p_2 \cdot p_3-D_{0}^{1,k} m_t^2
    \right.\\
&+& (2D_{13}^{1,k}+2D_{25}^{1,k}-D_{0}^{1,k} -
    D_{21}^{1,k}-2D_{11}^{1,k}) m_{\tilde{\chi}_1^+}^2+2(
    D_{311}^{1,k}-D_{313}^{1,k}
    +D_{313}^{3,k}-D_{27}^{3,k}-D_{311}^{3,k}\\
&-& \left. \left. D_{313}^{5,k}) \right] \right\}
    +(t \rightarrow b)
\end{eqnarray*}
\begin{eqnarray*}
f_6^{b,\hat{t}} &=&
\frac{i g_s^2}{16 \pi^2} \sum_{k=1,2} \left\{ 2D_{13}^{1,k} m_t
    m_{\tilde{\chi}_1^+} F_{1,k}+F_{2,k} \left[ 2(D_{37}^{1,k}-D_{33}^{1,k})
    \right. \right.
    p_1 \cdot p_2+2(D_{25}^{1,k}+D_{310}^{1,k}+D_{33}^{1,k}\\
&-& D_{23}^{1,k}-
    D_{37}^{1,k}-D_{39}^{1,k})
    p_1 \cdot p_3+2(D_{33}^{1,k}-D_{39}^{1,k}) p_2 \cdot p_3+D_{13}^{1,k} m_t^2
    +(D_{13}^{1,k}+2D_{37}^{1,k}-D_{35}^{1,k}\\
&-& \left. \left. 2D_{33}^{1,k}) m_{\tilde{\chi}_1^+}^2 +
    4D_{313}^{1,k}+2D_{313}^{3,k}+2D_{312}^{5,k}-2D_{313}^{5,k} \right]
     \right\}
    +(t \rightarrow b)
\end{eqnarray*}
\begin{eqnarray*}
f_7^{b,\hat{t}} &=&
\frac{i g_s^2}{8 \pi^2} \sum_{k=1,2} \left[ m_t F_{1,k} (D_{12}^{1,k}+
    D_{24}^{1,k}+D_{25}^{1,k}+D_{12}^{3,k}+D_{24}^{3,k}+D_{25}^{3,k} \right.
    +D_{25}^{5,k}-D_{11}^{1,k}-D_{21}^{1,k}\\
&-& D_{26}^{1,k}- D_{11}^{3,k}-
    D_{21}^{3,k}-D_{26}^{3,k}-D_{26}^{5,k}) +
    m_{\tilde{\chi}_1^+} F_{2,k} (2D_{24}^{1,k}+D_{12}^{1,k}+D_{25}^{1,k}+
    D_{34}^{1,k}+D_{35}^{1,k}\\
&+& D_{21}^{3,k}+D_{310}^{3,k}+
    D_{31}^{3,k}+D_{38}^{5,k}-2D_{21}^{1,k}-D_{11}^{1,k}-D_{26}^{1,k}-
    D_{310}^{1,k}-D_{31}^{1,k}-D_{24}^{3,k}-D_{34}^{3,k}\\
&-& \left. D_{35}^{3,k}-D_{310}^{5,k}) \right]
    +(t \rightarrow b)
\end{eqnarray*}
\begin{eqnarray*}
f_8^{b,\hat{t}} &=&
\frac{i g_s^2}{8 \pi^2} \sum_{k=1,2} \left[ m_t F_{1,k} (D_{12}^{1,k}+
    D_{24}^{1,k}+D_{0}^{3,k}+D_{11}^{3,k}+D_{12}^{3,k}+D_{24}^{3,k} \right.
    + D_{13}^{5,k}+D_{25}^{5,k}
    - D_{26}^{1,k}\\
&-& D_{13}^{3,k}-D_{26}^{3,k})+
    m_{\tilde{\chi}_1^+} F_{2,k} (2D_{24}^{1,k}+D_{12}^{1,k}
    + D_{34}^{1,k}+D_{25}^{3,k}+D_{310}^{3,k}-D_{26}^{1,k}-D_{310}^{1,k}-
    D_{11}^{3,k}\\
&-& D_{21}^{3,k}-D_{24}^{3,k}-D_{34}^{3,k}
    - \left. D_{26}^{5,k}-D_{310}^{5,k}) \right]
    +(t \rightarrow b)
\end{eqnarray*}
\begin{eqnarray*}
f_9^{b,\hat{t}} &=&
\frac{i g_s^2}{8 \pi^2} \sum_{k=1,2} \left[ m_t F_{1,k} (D_{25}^{1,k}+
    D_{25}^{3,k}+D_{22}^{5,k}+D_{25}^{5,k}-D_{26}^{1,k}-D_{26}^{3,k} \right.
    -D_{24}^{5,k}-D_{26}^{5,k})\\
&+& m_{\tilde{\chi}_1^+} F_{2,k} (D_{25}^{1,k}+
    D_{35}^{1,k}+D_{310}^{3,k}+D_{36}^{5,k}+D_{38}^{5,k}
     -\left. D_{26}^{1,k}-D_{310}^{1,k}-D_{35}^{3,k}-D_{310}^{5,k}-D_{32}^{5,k})
     \right] \\
&+& (t \rightarrow b)
\end{eqnarray*}
\begin{eqnarray*}
f_{10}^{b,\hat{t}} &=&
\frac{i g_s^2}{8 \pi^2} \sum_{k=1,2} \left[ m_t F_{1,k} (D_{13}^{5,k}+
    D_{25}^{5,k}-D_{26}^{1,k}-D_{13}^{3,k}-D_{26}^{3,k}-D_{12}^{5,k} \right.
    -D_{24}^{5,k})\\
&+& m_{\tilde{\chi}_1^+} F_{2,k} (D_{25}^{3,k}+D_{310}^{3,k}+
    D_{22}^{5,k}+D_{36}^{5,k}-D_{26}^{1,k}-D_{310}^{1,k}
    - \left. D_{26}^{5,k}-D_{310}^{5,k}) \right]
    +(t \rightarrow b)
\end{eqnarray*}
\begin{eqnarray*}
f_{11}^{b,\hat{t}} &=&
\frac{i g_s^2}{32 \pi^2} \sum_{k=1,2} (-2D_{0}^{1,k} m_t m_{\tilde{\chi}_1^+}
    F_{1,k}+F_{2,k} \left\{ 2(D_{26}^{1,k}+D_{310}^{1,k}+D_{33}^{1,k} \right.
    - D_{25}^{1,k}-D_{37}^{1,k}\\
&-& D_{39}^{1,k}) p_1 \cdot p_2
    +2(D_{22}^{1,k}+
    D_{23}^{1,k}+D_{36}^{1,k}+D_{37}^{1,k}+2D_{39}^{1,k}
    -D_{33}^{1,k}-D_{38}^{1,k}-2D_{26}^{1,k}\\
&-& 2D_{310}^{1,k}) p_1 \cdot p_3+2(2
    D_{39}^{1,k}-D_{33}^{1,k}-D_{38}^{1,k})
    p_2 \cdot p_3+(D_{12}^{1,k}-D_{0}^{1,k}-D_{13}^{1,k}) m_t^2\\
&+& \left[ 2(
    D_{26}^{1,k}+D_{310}^{1,k}+D_{33}^{1,k}-D_{24}^{1,k} \right.
    -  D_{37}^{1,k}-D_{39}^{1,k})+D_{13}^{1,k}+D_{21}^{1,k}+D_{35}^{1,k}-
    D_{0}^{1,k}\\
&-& D_{12}^{1,k} \left. -D_{34}^{1,k} \right]
    m_{\tilde{\chi}_1^+}^2 +
    4(D_{312}^{1,k}-D_{313}^{1,k})+2(D_{27}^{5,k}+D_{311}^{5,k}\\
&+& D_{313}^{5,k}+
   \left. D_{312}^{3,k}-D_{313}^{3,k}-D_{312}^{5,k}) \right\})
    +(t \rightarrow b)
\end{eqnarray*}
\begin{eqnarray*}
f_{12}^{b,\hat{t}} &=&
\frac{i g_s^2}{16 \pi^2} \sum_{k=1,2} F_{2,k} (D_{313}^{1,k}+D_{311}^{5,k}+
    D_{313}^{5,k}+D_{312}^{3,k}-D_{27}^{1,k}-D_{312}^{1,k}-D_{313}^{3,k}
    - D_{312}^{5,k})
    +(t \rightarrow b)
\end{eqnarray*}
\begin{eqnarray*}
f_{13}^{b,\hat{t}} &=&
\frac{i g_s^2}{16 \pi^2} \sum_{k=1,2} \left[ m_t F_{1,k} (D_{12}^{1,k}+
    D_{12}^{5,k}-D_{11}^{1,k}-D_{11}^{5,k}) \right.\\
&+& \left. m_{\tilde{\chi}_1^+} F_{2,k} (D_{12}^{1,k}+D_{24}^{1,k}+D_{24}^{5,k}
    -D_{11}^{1,k}-D_{21}^{1,k}-D_{22}^{5,k}) \right]
    +(t \rightarrow b)
\end{eqnarray*}
\begin{eqnarray*}
f_{14}^{b,\hat{t}} &=&
\frac{i g_s^2}{16 \pi^2} \sum_{k=1,2} \left[ m_t F_{1,k} (D_{12}^{1,k}-
    D_{0}^{5,k}-D_{11}^{5,k})+m_{\tilde{\chi}_1^+} F_{2,k} (D_{12}^{1,k}
     \right.
    + \left. D_{24}^{1,k}+D_{12}^{5,k}+D_{24}^{5,k}) \right]\\
&+& (t \rightarrow b)
\end{eqnarray*}
\begin{eqnarray*}
f_{15}^{b,\hat{t}} &=&
\frac{i g_s^2}{16 \pi^2} \sum_{k=1,2} \left[ m_t F_{1,k} (D_{11}^{1,k}-
    D_{13}^{1,k})+m_{\tilde{\chi}_1^+} F_{2,k} (D_{11}^{1,k}+D_{21}^{1,k}
     \right.
    - \left. D_{13}^{1,k}-D_{25}^{1,k}) \right]
    +(t \rightarrow b)
\end{eqnarray*}
\begin{eqnarray*}
f_{16}^{b,\hat{t}} &=&
-\frac{i g_s^2}{16 \pi^2} \sum_{k=1,2} \left[ D_{13}^{1,k} m_t F_{1,k}+
    m_{\tilde{\chi}_1^+} F_{2,k} (D_{13}^{1,k}+D_{25}^{1,k}) \right]
    +(t \rightarrow b)
\end{eqnarray*}
\begin{eqnarray*}
f_{17}^{b,\hat{t}} &=&
\frac{i g_s^2}{8 \pi^2} \sum_{k=1,2} F_{2,k} (D_{24}^{1,k}+D_{26}^{1,k}+
    D_{34}^{1,k}+D_{37}^{1,k}+D_{38}^{1,k}+D_{22}^{3,k}+D_{25}^{3,k}
    + D_{35}^{3,k}+D_{36}^{3,k}\\
&+& D_{39}^{3,k}+D_{25}^{5,k}+D_{35}^{5,k}+
    D_{37}^{5,k}+D_{38}^{5,k}-D_{22}^{1,k}-D_{25}^{1,k}-
    D_{35}^{1,k}-D_{36}^{1,k}-D_{39}^{1,k}-D_{24}^{3,k}\\
&-& D_{26}^{3,k}-
    D_{34}^{3,k}-D_{37}^{3,k}-D_{38}^{3,k}-D_{26}^{5,k}
    - D_{39}^{5,k}-2D_{310}^{5,k})
    +(t \rightarrow b)
\end{eqnarray*}
\begin{eqnarray*}
f_{18}^{b,\hat{t}} &=&
\frac{i g_s^2}{8 \pi^2} \sum_{k=1,2} F_{2,k} (D_{25}^{1,k}+D_{26}^{1,k}+
    D_{310}^{1,k}+D_{38}^{1,k}+D_{12}^{3,k}+D_{22}^{3,k}+D_{23}^{3,k}
    +D_{24}^{3,k}+D_{36}^{3,k}\\
&+& D_{39}^{3,k}+2D_{25}^{5,k}+D_{13}^{5,k}+
    D_{23}^{5,k}+D_{35}^{5,k}+D_{37}^{5,k}-D_{22}^{1,k}
    -D_{23}^{1,k}-D_{36}^{1,k}-D_{39}^{1,k}-2D_{26}^{3,k}\\
&-& D_{13}^{3,k}-
    D_{25}^{3,k}-D_{310}^{3,k}-D_{38}^{3,k}-D_{26}^{5,k}
    -D_{310}^{5,k})
    +(t \rightarrow b)
\end{eqnarray*}
\begin{eqnarray*}
f_{19}^{b,\hat{t}} &=&
\frac{i g_s^2}{8 \pi^2} \sum_{k=1,2} F_{2,k} (D_{26}^{1,k}+D_{37}^{1,k}+
    D_{38}^{1,k}+D_{310}^{3,k}+D_{39}^{3,k}+2D_{36}^{5,k}+2D_{38}^{5,k}
    +D_{35}^{5,k}+D_{37}^{5,k}\\
&-& D_{25}^{1,k}-D_{310}^{1,k}-D_{39}^{1,k}-
    D_{37}^{3,k}-
    D_{38}^{3,k}-3D_{310}^{5,k}-D_{32}^{5,k}- D_{34}^{5,k}-D_{39}^{5,k})
    +(t \rightarrow b)
\end{eqnarray*}
\begin{eqnarray*}
f_{20}^{b,\hat{t}} &=&
\frac{i g_s^2}{8 \pi^2} \sum_{k=1,2} F_{2,k} (D_{26}^{1,k}+D_{38}^{1,k}+
    D_{23}^{3,k}+D_{39}^{3,k}+D_{22}^{5,k}+D_{23}^{5,k}+D_{25}^{5,k}
    +D_{35}^{5,k}+D_{36}^{5,k}\\
&+& D_{37}^{5,k}-D_{24}^{5,k}-D_{23}^{1,k}-
    D_{39}^{1,k}-D_{26}^{3,k}-D_{38}^{3,k}-D_{34}^{5,k}-
    2D_{26}^{5,k}-2D_{310}^{5,k})
    +(t \rightarrow b)
\end{eqnarray*}
\begin{eqnarray*}
f_{34}^{b,\hat{t}} &=&
\frac{i g_s^2}{32 \pi^2} \sum_{k=1,2} (m_t F_{3,k} \left\{ 2(D_{13}^{1,k}+
    D_{25}^{1,k}-D_{23}^{1,k}) p_1 \cdot p_2+2(D_{11}^{1,k}+D_{12}^{1,k}
     \right.+
    D_{23}^{1,k}+D_{24}^{1,k}\\
&-& D_{25}^{1,k} -D_{26}^{1,k}-2D_{13}^{1,k})
    p_1 \cdot p_3+2(D_{23}^{1,k}-D_{13}^{1,k}-D_{26}^{1,k})
    p_2 \cdot p_3+\left[ 2(D_{13}^{1,k}+D_{25}^{1,k} \right.\\
&-& \left. D_{11}^{1,k}-D_{23}^{1,k})
    -D_{0}^{1,k}-D_{21}^{1,k} \right]
    \left. m_{\tilde{\chi}_1^+}^2+4D_{27}^{1,k}-2(D_{27}^{5,k}+D_{27}^{1,k}+
    D_{27}^{3,k})+D_{0}^{1,k} m_t^2 \right\}\\
&+& m_{\tilde{\chi}_1^+} F_{4,k} \left[ 2(D_{35}^{1,k}+2D_{33}^{1,k}-3
    D_{37}^{1,k}-D_{13}^{1,k}-D_{23}^{1,k})   \right.
    p_1 \cdot p_2+2(3D_{23}^{1,k}+3D_{37}^{1,k}
    + D_{11}^{1,k}\\
&+& D_{21}^{1,k}+
    D_{34}^{1,k}+2D_{39}^{1,k}-3D_{25}^{1,k}-3D_{310}^{1,k}
    - D_{12}^{1,k}-D_{26}^{1,k}-D_{35}^{1,k}-2D_{33}^{1,k}) p_1 \cdot p_3\\
&+& 2(
    D_{13}^{1,k}+D_{23}^{1,k}+D_{26}^{1,k}+D_{37}^{1,k}
    +2D_{39}^{1,k}-D_{25}^{1,k}-D_{310}^{1,k}-2D_{33}^{1,k}) p_2 \cdot p_3
    +(D_{11}^{1,k}-2D_{13}^{1,k}\\
&-& D_{0}^{1,k}) m_t^2 +4(D_{311}^{1,k}+D_{313}^{5,k}
    -2D_{313}^{1,k}-D_{313}^{3,k}) +2(D_{311}^{3,k}-D_{312}^{5,k})\\
&+& (D_{0}^{1,k}+D_{11}^{1,k}+4D_{25}^{1,k}+4
    D_{33}^{1,k}+4D_{35}^{1,k}-6D_{37}^{1,k}-2D_{23}^{1,k}
    - \left. D_{21}^{1,k}-D_{31}^{1,k}) m_{\tilde{\chi}_1^+}^2 \right])\\
&+& (t \rightarrow b)
\end{eqnarray*}
\begin{eqnarray*}
f_{35}^{b,\hat{t}} &=&
\frac{i g_s^2}{16 \pi^2} \sum_{k=1,2} \left\{ -m_t F_{3,k} (D_{27}^{1,k}+
    D_{27}^{3,k}+D_{27}^{5,k})+m_{\tilde{\chi}_1^+} F_{4,k} \left[ 2(
    D_{313}^{1,k} \right. \right.
    + D_{313}^{5,k}-D_{313}^{3,k})\\
&+& D_{311}^{3,k}-D_{27}^{1,k}-
    \left. \left.D_{311}^{1,k}-D_{312}^{5,k} \right] \right\}
    +(t \rightarrow b)
\end{eqnarray*}
\begin{eqnarray*}
f_{36}^{b,\hat{t}} &=&
\frac{i g_s^2}{16 \pi^2} \sum_{k=1,2} F_{4,k} \left\{ 2(D_{26}^{1,k}+
    D_{310}^{1,k}+D_{37}^{1,k}+D_{37}^{5,k}+D_{38}^{5,k}-D_{25}^{1,k}-
    D_{35}^{1,k} \right.
    - D_{39}^{1,k}-D_{310}^{5,k}\\
&-& D_{39}^{5,k}) p_1 \cdot p_2+2(D_{22}^{1,k}+
    D_{25}^{1,k}+D_{35}^{1,k}+D_{36}^{1,k}+D_{39}^{1,k}
    +2D_{310}^{5,k}+D_{26}^{5,k}+D_{39}^{5,k}-D_{24}^{1,k}\\
&-& D_{26}^{1,k}-
    D_{34}^{1,k}-D_{37}^{1,k}-D_{38}^{1,k}-D_{25}^{5,k}-
    D_{35}^{5,k}-D_{37}^{5,k}-D_{38}^{5,k}) p_1 \cdot p_3+2(D_{25}^{1,k}+
    D_{310}^{1,k}\\
&+& D_{39}^{1,k}+3D_{310}^{5,k}+D_{32}^{5,k}
    + D_{34}^{5,k}+D_{39}^{5,k}-D_{26}^{1,k}-D_{37}^{1,k}-D_{38}^{1,k}-2
    D_{36}^{5,k}-2D_{38}^{5,k}-D_{35}^{5,k}\\
&-& D_{37}^{5,k})
    p_2 \cdot p_3+(D_{12}^{1,k}+D_{12}^{5,k}-D_{11}^{1,k}-D_{11}^{5,k}) m_t^2+
    \left[ D_{11}^{1,k}+D_{31}^{1,k}-D_{12}^{1,k} \right.
    - D_{34}^{1,k}+D_{36}^{5,k}\\
&-& D_{32}^{5,k}+2(D_{21}^{1,k}+D_{26}^{1,k}+
    D_{310}^{1,k}+D_{37}^{1,k}+D_{37}^{5,k}+D_{38}^{5,k}
    - D_{24}^{1,k}-D_{25}^{1,k}-D_{35}^{1,k}-D_{39}^{1,k}\\
&-& D_{310}^{5,k}- \left. D_{39}^{5,k}) \right] m_{\tilde{\chi}_1^+}^2+
    4(D_{312}^{1,k}+D_{312}^{5,k}- D_{311}^{1,k} - D_{311}^{5,k}) \\
&+& \left. 2(D_{312}^{3,k}-D_{311}^{3,k}) \right\}
    +(t \rightarrow b)
\end{eqnarray*}
\begin{eqnarray*}
f_{37}^{b,\hat{t}} &=&
\frac{i g_s^2}{16 \pi^2} \sum_{k=1,2} F_{4,k} \left\{ 2(D_{26}^{1,k}+
    D_{310}^{1,k}+D_{33}^{1,k}+D_{23}^{5,k}+D_{37}^{5,k}-D_{23}^{1,k}-
    D_{37}^{1,k} \right.
    - D_{39}^{1,k}-D_{26}^{5,k}\\
&-& D_{310}^{5,k}) p_1 \cdot p_2+2\left[ 2(
    D_{23}^{1,k}+D_{39}^{1,k}-D_{26}^{1,k}-D_{310}^{1,k}) \right.
    + D_{22}^{1,k}+D_{36}^{1,k}+D_{37}^{1,k}+D_{26}^{5,k}
    + D_{310}^{5,k}\\
&-& D_{25}^{1,k}-D_{33}^{1,k}-D_{38}^{1,k}-D_{13}^{5,k}
    - \left. D_{23}^{5,k}-D_{35}^{5,k}-D_{37}^{5,k}-2D_{25}^{5,k} \right]
    p_1 \cdot p_3+2(2D_{39}^{1,k}+D_{23}^{1,k}\\
&+& 2D_{26}^{5,k}+2D_{310}^{5,k}
    + D_{24}^{5,k}+D_{34}^{5,k}-D_{26}^{1,k}-D_{33}^{1,k}-D_{38}^{1,k}-
    D_{22}^{5,k}-D_{23}^{5,k}-D_{25}^{5,k}-D_{35}^{5,k}\\
&-& D_{36}^{5,k}-D_{37}^{5,k}) p_2 \cdot p_3+(D_{12}^{1,k}-D_{13}^{1,k}-
    D_{0}^{5,k}-D_{11}^{5,k})
    m_t^2+\left[ 2(D_{25}^{1,k}+D_{26}^{1,k}+D_{310}^{1,k}+D_{33}^{1,k}\right.\\
&+& D_{23}^{5,k}+D_{37}^{5,k}-D_{23}^{1,k}-D_{24}^{1,k}
    - D_{37}^{1,k}-D_{39}^{1,k}-D_{26}^{5,k}-D_{310}^{5,k})+D_{13}^{1,k}+
    D_{35}^{1,k}+D_{22}^{5,k}\\
&+& D_{36}^{5,k}-D_{12}^{1,k}
    - \left. D_{34}^{1,k} \right] m_{\tilde{\chi}_1^+}^2 -
    6D_{313}^{1,k}+4(D_{312}^{1,k}- D_{311}^{5,k})\\
&+& \left. 2(D_{27}^{1,k}+D_{27}^{3,k}+D_{312}^{3,k}-D_{27}^{5,k}) \right\}
    +(t \rightarrow b)
\end{eqnarray*}
\begin{eqnarray*}
f_{38}^{b,\hat{t}} &=&
\frac{i g_s^2}{16 \pi^2} \sum_{k=1,2} F_{4,k} \left[ 2(D_{26}^{1,k}-
    D_{25}^{1,k}) p_2 \cdot p_3-D_{0}^{1,k} m_t^2+(D_{0}^{1,k}+D_{21}^{1,k}
     \right.
    + 2D_{11}^{1,k}) m_{\tilde{\chi}_1^+}^2\\
&+& \left. 2(D_{311}^{1,k}+D_{313}^{3,k}-
    D_{313}^{1,k}-D_{27}^{3,k}-D_{311}^{3,k}-D_{313}^{5,k}) \right]
    +(t \rightarrow b)
\end{eqnarray*}
\begin{eqnarray*}
f_{39}^{b,\hat{t}} &=&
\frac{i g_s^2}{16 \pi^2} \sum_{k=1,2} F_{4,k} \left[ 2(D_{37}^{1,k}-
    D_{33}^{1,k}) p_1 \cdot p_2+2(D_{25}^{1,k}+D_{310}^{1,k}+D_{33}^{1,k}-
    D_{23}^{1,k} \right.
    - D_{37}^{1,k}\\
&-& D_{39}^{1,k}) p_1 \cdot p_3
    + 2(D_{33}^{1,k}-D_{39}^{1,k})
    p_2 \cdot p_3+D_{13}^{1,k} m_t^2+(2D_{37}^{1,k}
    - 2D_{25}^{1,k}-2D_{33}^{1,k}-D_{13}^{1,k}\\
&-& D_{35}^{1,k}) m_{\tilde{\chi}_1^+} ^2
    + 4D_{313}^{1,k}+2(D_{313}^{3,k}
    + \left. D_{312}^{5,k}-D_{313}^{5,k}) \right]
    +(t \rightarrow b)
\end{eqnarray*}
\begin{eqnarray*}
f_{40}^{b,\hat{t}} &=&
\frac{i g_s^2}{8 \pi^2} \sum_{k=1,2} \left\{ m_t F_{3,k} (D_{11}^{1,k}+
    D_{21}^{1,k}+D_{26}^{1,k}+D_{11}^{3,k}+D_{21}^{3,k}+D_{26}^{3,k} \right.
    + D_{26}^{5,k}-D_{12}^{1,k}-D_{24}^{1,k}\\
&-& D_{25}^{1,k}-D_{12}^{3,k}-
    D_{24}^{3,k}-D_{25}^{3,k}-D_{25}^{5,k})
    + m_{\tilde{\chi}_1^+} F_{4,k} \left[ 3(D_{26}^{1,k}+D_{310}^{1,k}+
    D_{35}^{3,k}-D_{25}^{1,k}-D_{35}^{1,k}\right. \\
&-& D_{310}^{3,k})
    + 2(D_{21}^{1,k}+D_{37}^{1,k}+D_{25}^{3,k}+D_{39}^{3,k}+D_{37}^{5,k}-
    D_{24}^{1,k}-D_{39}^{1,k}-D_{26}^{3,k}-D_{37}^{3,k}
    - D_{39}^{5,k})\\
&+& D_{11}^{1,k}+D_{31}^{1,k}+D_{24}^{3,k}+D_{34}^{3,k}+
    D_{38}^{5,k}-D_{12}^{1,k}-D_{34}^{1,k}-D_{21}^{3,k}
    - \left. \left. D_{31}^{3,k}-D_{310}^{5,k} \right] \right\}
    +(t \rightarrow b)
\end{eqnarray*}
\begin{eqnarray*}
f_{41}^{b,\hat{t}} &=&
\frac{i g_s^2}{8 \pi^2} \sum_{k=1,2} \left\{ m_t F_{3,k} (D_{26}^{1,k}+
    D_{13}^{3,k}+D_{26}^{3,k}-D_{12}^{1,k}-D_{24}^{1,k}-D_{0}^{3,k} \right.
    -D_{11}^{3,k}-D_{12}^{3,k}-D_{24}^{3,k}\\
&-& D_{13}^{5,k}-D_{25}^{5,k})+
    m_{\tilde{\chi}_1^+} F_{4,k} \left[ 3(D_{26}^{1,k}+D_{310}^{1,k} \right.
    - D_{25}^{3,k}-D_{310}^{3,k})+2(D_{13}^{1,k}+D_{25}^{1,k}+D_{23}^{3,k}\\
&+& D_{39}^{3,k} + D_{23}^{5,k}+D_{37}^{5,k}-D_{23}^{1,k}
    - D_{24}^{1,k}-D_{39}^{1,k}-D_{13}^{3,k}-D_{26}^{3,k})+D_{11}^{3,k}+
    D_{21}^{3,k}+D_{24}^{3,k}\\
&+& D_{34}^{3,k}-D_{12}^{1,k}
    - \left. \left. D_{34}^{1,k}-D_{26}^{5,k}-D_{310}^{5,k} \right] \right\}
    +(t \rightarrow b)
\end{eqnarray*}
\begin{eqnarray*}
f_{42}^{b,\hat{t}} &=&
\frac{i g_s^2}{8 \pi^2} \sum_{k=1,2} \left\{ m_t F_{3,k} (D_{26}^{1,k}+
    D_{26}^{3,k}+D_{24}^{5,k}+D_{26}^{5,k}-D_{25}^{1,k}-D_{25}^{3,k} \right.
    -D_{22}^{5,k}-D_{25}^{5,k})\\
&+& m_{\tilde{\chi}_1^+} F_{4,k} \left[ 3(
    D_{38}^{5,k}-D_{310}^{5,k})+2(D_{37}^{1,k}+D_{39}^{3,k}+D_{37}^{5,k}
     \right.
    - D_{39}^{1,k}-D_{37}^{3,k}-D_{39}^{5,k})+D_{26}^{1,k}\\
&+& D_{310}^{1,k}+ D_{35}^{3,k}+D_{36}^{5,k}-D_{25}^{1,k}-D_{35}^{1,k}
    - \left. \left. D_{310}^{3,k}-D_{32}^{5,k} \right] \right\}
    +(t \rightarrow b)
\end{eqnarray*}
\begin{eqnarray*}
f_{43}^{b,\hat{t}} &=&
\frac{i g_s^2}{8 \pi^2} \sum_{k=1,2} \left\{ m_t F_{3,k} (D_{26}^{1,k}+
    D_{13}^{3,k}+D_{26}^{3,k}+D_{12}^{5,k}+D_{24}^{5,k}-D_{13}^{5,k} \right.
    - D_{25}^{5,k})+m_{\tilde{\chi}_1^+} F_{4,k} \left[ D_{26}^{1,k}\right.\\
&+& D_{310}^{1,k}+D_{22}^{5,k}+D_{36}^{5,k}-D_{25}^{3,k}-D_{310}^{3,k}
    + 2(D_{23}^{3,k}+D_{39}^{3,k}+D_{23}^{5,k}+D_{37}^{5,k}-
    D_{23}^{1,k}-D_{39}^{1,k})\\
&-& \left. \left. 3(D_{26}^{5,k}+D_{310}^{5,k}) \right] \right\}
    +(t \rightarrow b)
\end{eqnarray*}
\begin{eqnarray*}
f_{44}^{b,\hat{t}} &=&
\frac{i g_s^2}{32 \pi^2} \sum_{k=1,2} F_{4,k} \left\{ 2(D_{26}^{1,k}+
    D_{310}^{1,k}+D_{33}^{1,k}-D_{25}^{1,k}-D_{37}^{1,k}-D_{39}^{1,k})
     \right.
    p_1 \cdot p_2+2(D_{22}^{1,k}+D_{23}^{1,k}\\
&+& D_{36}^{1,k}+D_{37}^{1,k}-
    D_{33}^{1,k}-D_{38}^{1,k}+2D_{39}^{1,k}-2D_{26}^{1,k}
    - 2D_{310}^{1,k}) p_1 \cdot p_3+2(2D_{39}^{1,k}-D_{33}^{1,k}\\
&-& D_{38}^{1,k}) p_2 \cdot p_3
    + (D_{12}^{1,k}-D_{0}^{1,k}-D_{13}^{1,k})
    m_t^2+\left[ 2(D_{11}^{1,k}+D_{26}^{1,k}+D_{310}^{1,k}+D_{33}^{1,k}-
    D_{24}^{1,k}\right.
    - D_{37}^{1,k}\\
&-& D_{39}^{1,k}) + \left. D_{0}^{1,k}+D_{13}^{1,k}+D_{21}^{1,k}+D_{35}^{1,k}
    -D_{12}^{1,k}-
    D_{34}^{1,k} \right] m_{\tilde{\chi}_1^+}^2 + 4(D_{312}^{1,k}\\
&-& \left. D_{313}^{1,k})+2(D_{27}^{5,k}+D_{311}^{5,k}
    +D_{313}^{5,k}+D_{312}^{3,k}-D_{313}^{3,k}-D_{312}^{5,k}) \right\}
    +(t \rightarrow b)
\end{eqnarray*}
\begin{eqnarray*}
f_{45}^{b,\hat{t}} &=&
\frac{i g_s^2}{16 \pi^2} \sum_{k=1,2} F_{4,k} (D_{313}^{1,k}+D_{312}^{3,k}+
    D_{311}^{5,k}+D_{313}^{5,k}-D_{27}^{1,k}-D_{312}^{1,k}-D_{313}^{3,k}-
    D_{312}^{5,k})
    +(t \rightarrow b)
\end{eqnarray*}
\begin{eqnarray*}
f_{46}^{b,\hat{t}} &=&
\frac{i g_s^2}{16 \pi^2} \sum_{k=1,2} \left\{ m_t F_{3,k} (D_{11}^{1,k}+
    D_{11}^{5,k}-D_{12}^{1,k}-D_{12}^{5,k}) \right.
    + m_{\tilde{\chi}_1^+} F_{4,k} \left[ 2(D_{26}^{1,k}+D_{26}^{5,k}-
    D_{25}^{1,k}\right.\\
&-&  D_{25}^{5,k})+ D_{11}^{1,k}+D_{21}^{1,k}+D_{24}^{5,k} -
    \left. \left. D_{12}^{1,k}-D_{24}^{1,k}-D_{22}^{5,k} \right] \right\}
    +(t \rightarrow b)
\end{eqnarray*}
\begin{eqnarray*}
f_{47}^{b,\hat{t}} &=&
\frac{i g_s^2}{16 \pi^2} \sum_{k=1,2} \left\{ m_t F_{3,k} (D_{0}^{5,k}+
    D_{11}^{5,k}-D_{12}^{1,k})+m_{\tilde{\chi}_1^+} F_{4,k} \left[ 2(
    D_{13}^{1,k} \right.\right.
    + D_{26}^{1,k}-D_{13}^{5,k}-D_{25}^{5,k})\\
&+& D_{12}^{5,k}
    + \left. \left. D_{24}^{5,k}-D_{12}^{1,k}-D_{24}^{1,k} \right] \right\}
    +(t \rightarrow b)
\end{eqnarray*}
\begin{eqnarray*}
f_{48}^{b,\hat{t}} &=&
\frac{i g_s^2}{16 \pi^2} \sum_{k=1,2} \left\{ m_t F_{3,k} (D_{13}^{1,k}-
    D_{11}^{1,k})+m_{\tilde{\chi}_1^+} F_{4,k} \left[ 2(D_{12}^{1,k}+
    D_{24}^{1,k} \right. \right.
    - D_{26}^{1,k})+D_{25}^{1,k}-D_{11}^{1,k}\\
&-& \left. \left. D_{13}^{1,k}- D_{21}^{1,k} \right] \right\}
    +(t \rightarrow b)
\end{eqnarray*}
\begin{eqnarray*}
f_{49}^{b,\hat{t}} &=&
\frac{i g_s^2}{16 \pi^2} \sum_{k=1,2} \left[ D_{13}^{1,k} m_t F_{3,k}+
    m_{\tilde{\chi}_1^+} F_{4,k} (D_{25}^{1,k}-D_{13}^{1,k}-2D_{26}^{1,k})
     \right]
    +(t \rightarrow b)
\end{eqnarray*}
\begin{eqnarray*}
f_{50}^{b,\hat{t}} &=&
\frac{i g_s^2}{8 \pi^2} \sum_{k=1,2} F_{4,k} (D_{24}^{1,k}+D_{26}^{1,k}+
    D_{34}^{1,k}+D_{37}^{1,k}+D_{38}^{1,k}+D_{22}^{3,k}+D_{25}^{3,k}
    + D_{35}^{3,k}+D_{36}^{3,k}
    + D_{39}^{3,k}\\
&+& D_{25}^{5,k}+D_{35}^{5,k}+
    D_{37}^{5,k}+D_{38}^{5,k}-D_{22}^{1,k}-D_{25}^{1,k}
    - D_{35}^{1,k}-D_{36}^{1,k}-D_{39}^{1,k}-D_{24}^{3,k}
    - D_{26}^{3,k}- D_{34}^{3,k}\\
&-& D_{37}^{3,k}-D_{38}^{3,k}-D_{26}^{5,k}
    - D_{39}^{5,k}-2D_{310}^{5,k})
    +(t \rightarrow b)
\end{eqnarray*}
\begin{eqnarray*}
f_{51}^{b,\hat{t}} &=&
\frac{i g_s^2}{8 \pi^2} \sum_{k=1,2} F_{4,k} (D_{25}^{1,k}+D_{26}^{1,k}+
    D_{310}^{1,k}+D_{38}^{1,k}+D_{12}^{3,k}+D_{22}^{3,k}+D_{23}^{3,k}
    + D_{24}^{3,k}+D_{36}^{3,k}+D_{39}^{3,k}\\
&+& D_{13}^{5,k}+D_{23}^{5,k}+
    D_{35}^{5,k}+D_{37}^{5,k}+2D_{25}^{5,k}-D_{22}^{1,k}
    - D_{23}^{1,k}-D_{36}^{1,k}-D_{39}^{1,k}-2D_{26}^{3,k}-D_{13}^{3,k}-
    D_{25}^{3,k}\\
&-& D_{310}^{3,k}-D_{38}^{3,k}-D_{26}^{5,k}- D_{310}^{5,k})
    +(t \rightarrow b)
\end{eqnarray*}
\begin{eqnarray*}
f_{52}^{b,\hat{t}} &=&
\frac{i g_s^2}{8 \pi^2} \sum_{k=1,2} F_{4,k} (D_{26}^{1,k}+D_{37}^{1,k}+
    D_{38}^{1,k}+D_{310}^{3,k}+D_{39}^{3,k}+D_{35}^{5,k}+D_{37}^{5,k}
    + 2D_{36}^{5,k}+2D_{38}^{5,k}-D_{25}^{1,k}\\
&-& D_{310}^{1,k}-D_{39}^{1,k}- D_{37}^{3,k}-D_{38}^{3,k}-D_{32}^{5,k}
    -D_{34}^{5,k}- D_{39}^{5,k}-3D_{310}^{5,k})
    +(t \rightarrow b)
\end{eqnarray*}
\begin{eqnarray*}
f_{53}^{b,\hat{t}} &=&
\frac{i g_s^2}{8 \pi^2} \sum_{k=1,2} F_{4,k} (D_{26}^{1,k}+D_{38}^{1,k}+
    D_{23}^{3,k}+D_{39}^{3,k}+D_{22}^{5,k}+D_{23}^{5,k}+D_{25}^{5,k}
    +D_{35}^{5,k}+D_{36}^{5,k}+D_{37}^{5,k}\\
&-& D_{23}^{1,k}-D_{39}^{1,k}-
    D_{26}^{3,k}-D_{38}^{3,k}-D_{24}^{5,k}-D_{34}^{5,k}
    - 2D_{26}^{5,k}-2D_{310}^{5,k})
    +(t \rightarrow b)
\end{eqnarray*}
\begin{eqnarray*}
f_{i}^{b,\hat{t}} = 0 \hskip 5mm (i=21\sim 33, 54\sim 66)
\end{eqnarray*}
\par
In this work we adopted the definitions of two-, three-, four-point one-loop
Passarino-Veltman integral functions as shown in reference\cite{s19} and
all the vector and tensor integrals can be deduced in the forms of scalar
integrals \cite{s20}.

\begin{center}
{\large \bf Figure Captions}
\end{center}

\parindent=0pt

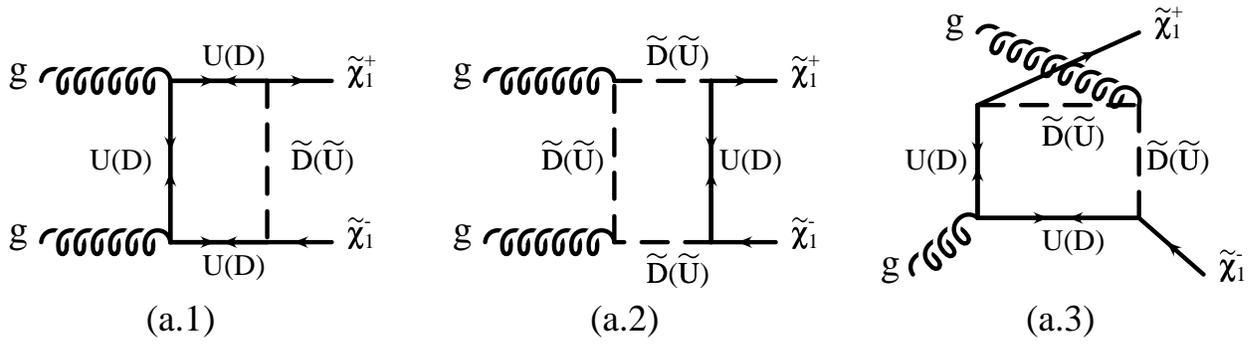
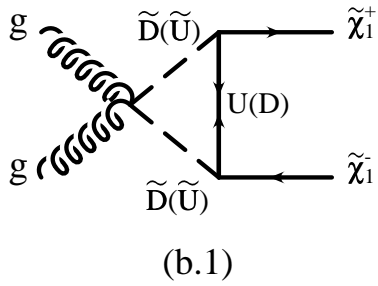
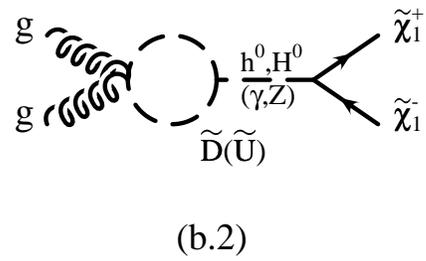
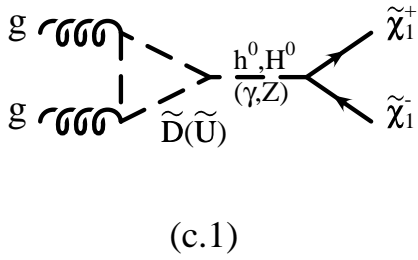
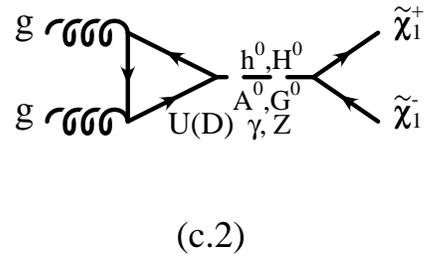
{\bf Fig.1} Feynman diagrams at one-loop level of the subprocess
     $gg \rightarrow \tilde{\chi}_1^{+} \tilde{\chi}_1^{-}$, where
     $(U,D)=(u,d),(c,s),(t,b)$. $(a.1 \sim a.3)$ box diagrams.
     $(b.1 \sim b.2)$ quartic interaction diagrams. $(c.1 \sim c.3)$
     triangle interaction s-channel diagrams. The figures with exchanging
     incoming gluons in Fig.1(a) and Fig.1(c) are not shown.

{\bf Fig.2(a)} The cross section $\hat{\sigma}$ of the higgsino-like chargino
     pair production subprocess $gg \rightarrow \tilde{\chi}_1^{+}
     \tilde{\chi}_1^{-}$ versus the c.m.s. energy of incoming gluons
     $\sqrt{\hat{s}}$ with $m_{\tilde{\chi}^{+}_{1}}=165 GeV$,
     $m_{\tilde{\chi}^{+}_{2}}=750 GeV$, $\tilde{M}=200 GeV$ (for the third
     generation), $m_{A}=250 GeV$, the masses of the first and second
     generation squarks being 600 GeV and all vanishing CP-odd phases.
     The full-line is for $\tan{\beta}=4$. The dashed-line for
     $\tan{\beta}=40$.

{\bf Fig.2(b)} The cross section $\hat{\sigma}$ of the gaugino-like chargino
     pair production subprocess $gg \rightarrow \tilde{\chi}_1^{+}
     \tilde{\chi}_1^{-}$ versus the c.m.s. energy of incoming gluons
     $\sqrt{\hat{s}}$ with $m_{\tilde{\chi}^{+}_{1}}=165 GeV$,
     $m_{\tilde{\chi}^{+}_{2}}=750 GeV$, $m_{A}=250 GeV$, $\tilde{M}=200 GeV$
     (for the third generation), the masses of the first and second
     generation squarks being 600 GeV and all vanishing CP-odd
     phases. The full-line is for $\tan{\beta}=4$ and the dashed-line for
     $\tan{\beta}=40$.

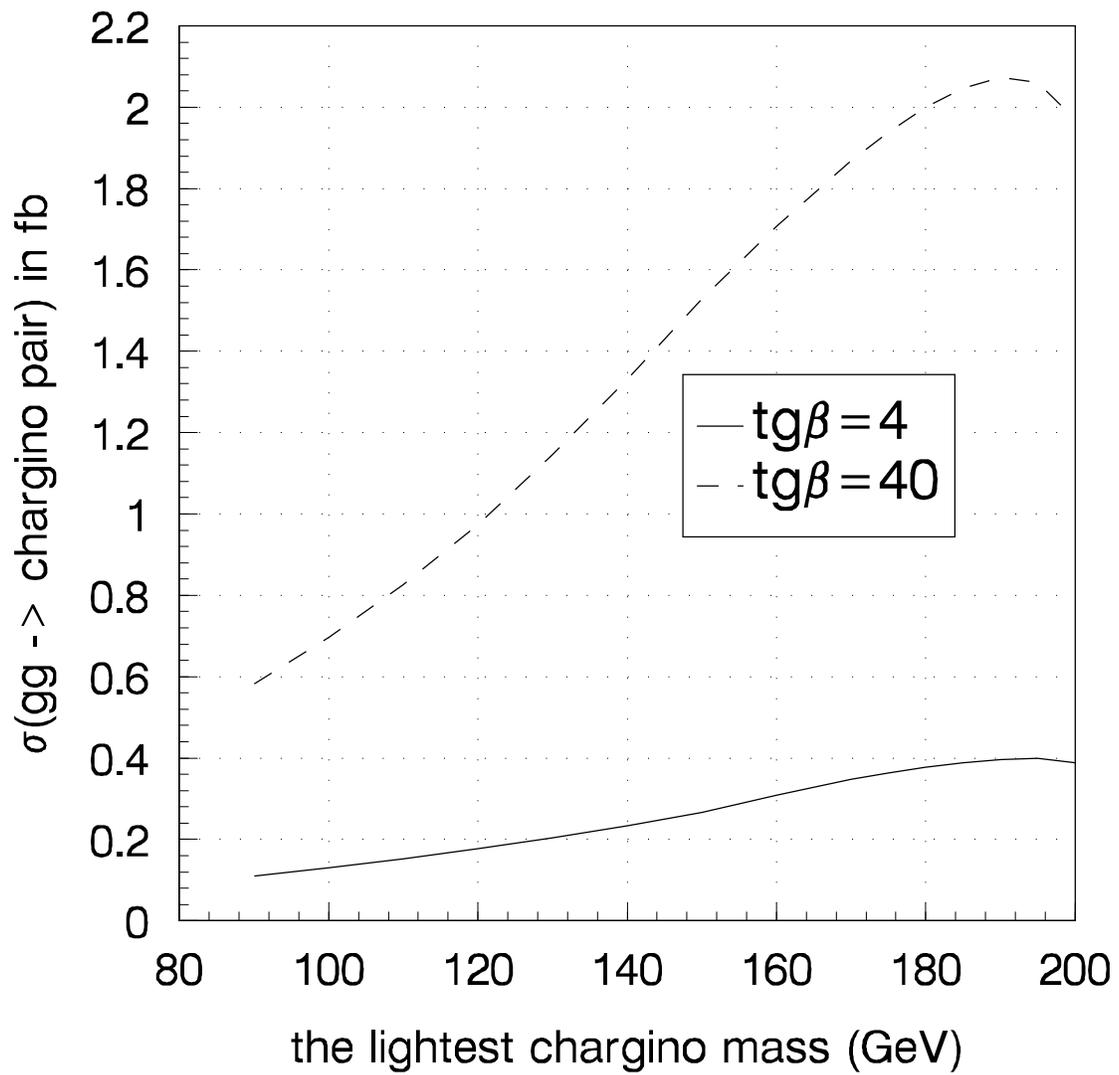
{\bf Fig.3} The cross section $\hat{\sigma}$ of the higgsino-like chargino
     pair production subprocess $gg \rightarrow \tilde{\chi}_1^{+}
     \tilde{\chi}_1^{-}$ versus the lightest chargino mass
     with $\sqrt{\hat{s}}=450 GeV$, $m_{\tilde{\chi}^{+}_{1}}=165 GeV$,
     $m_{\tilde{\chi}^{+}_{2}}=750 GeV$, $m_{A}=250 GeV$, $\tilde{M}=200 GeV$
     (for the third generation), the masses of the first and second
     generation squarks being 600 GeV and all vanishing CP-odd phases.
     The full-line is for $\tan{\beta}=4$ and the dashed-line for
     $\tan{\beta}=40$.

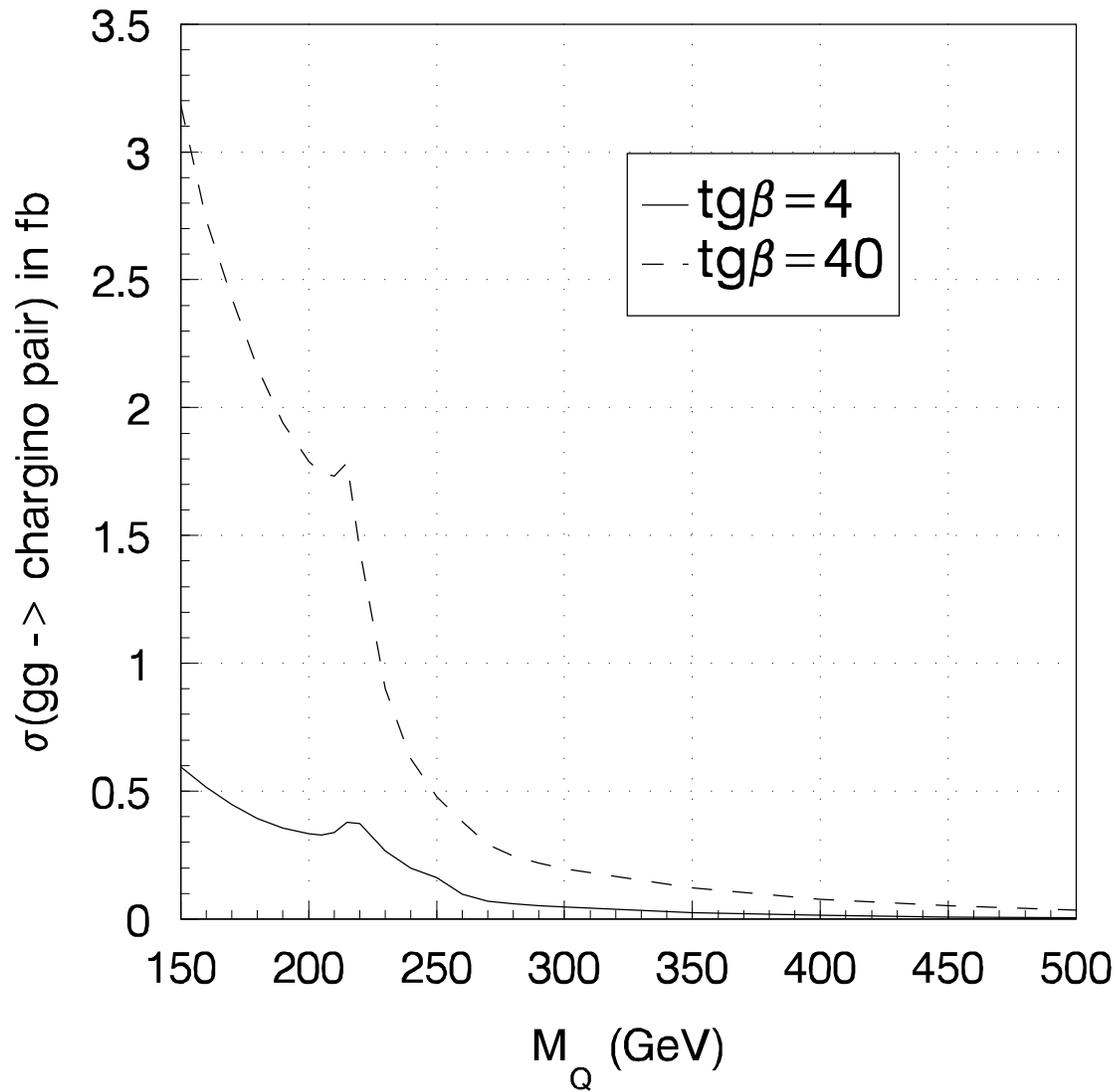
{\bf Fig.4} The cross section $\hat{\sigma}$ of the higgsino-like chargino
     pair production subprocess $gg \rightarrow \tilde{\chi}_1^{+}
     \tilde{\chi}_1^{-}$ versus the third generation $\tilde{M}
     (=M_{\tilde{Q}} = M_{\tilde{t}} = M_{\tilde{b}})$ with
     $\sqrt{\hat{s}}=450 GeV$, $m_{\tilde{\chi}^{+}_{1}}=165 GeV$,
     $m_{\tilde{\chi}^{+}_{2}}=750 GeV$, $\phi_{\mu}=\phi_{\tilde{U}}=
     \phi_{\tilde{D}}=0$, $m_{A}=250 GeV$ and all the masses of the first and
     second generation squarks being 600  GeV. The full-line is for
     $\tan{\beta}=4$ and the dashed-line for $\tan{\beta}=40$.

{\bf Fig.5(a)} The cross section $\hat{\sigma}$ of the higgsino-like chargino
     pair production subprocess $gg \rightarrow \tilde{\chi}_1^{+}
     \tilde{\chi}_1^{-}$ versus $\phi_{CP}$'s with $\sqrt{\hat{s}}=450 GeV$,
     $m_{\tilde{\chi}^{+}_{1}}=165 GeV$, $m_{\tilde{\chi}^{+}_{2}}=750 GeV$,
     $\tan{\beta}=4$, $m_{A}=250 GeV$, $\tilde{M}=200 GeV$ (for the third
     generation) and the masses of the first and second generation
     squarks being 600 GeV. The full-line is for $\hat{\sigma}$ versus
     $\phi_{q}(=\phi_{\tilde{t}} =\phi_{\tilde{b}})$. The dashed-line
     is for $\hat{\sigma}$ versus $\phi_{\mu}$.

{\bf Fig.5(b)} The cross section $\hat{\sigma}$ of the higgsino-like chargino
     pair production subprocess $gg \rightarrow \tilde{\chi}_1^{+}
     \tilde{\chi}_1^{-}$ versus $\phi_{CP}$'s with $\sqrt{\hat{s}}=450 GeV$,
     $m_{\tilde{\chi}^{+}_{1}}=165 GeV$, $m_{\tilde{\chi}^{+}_{2}}=750 GeV$,
     $\tan{\beta}=40$, $m_{A}=250 GeV$, $\tilde{M}=200 GeV$ (for the third
     generation) and the masses of the first and second generation
     squarks being 600 GeV. The full-line is for $\hat{\sigma}$ versus
     $\phi_{q}(=\phi_{\tilde{t}} =\phi_{\tilde{b}})$. The dashed-line is
     for $\hat{\sigma}$ versus $\phi_{\mu}$.

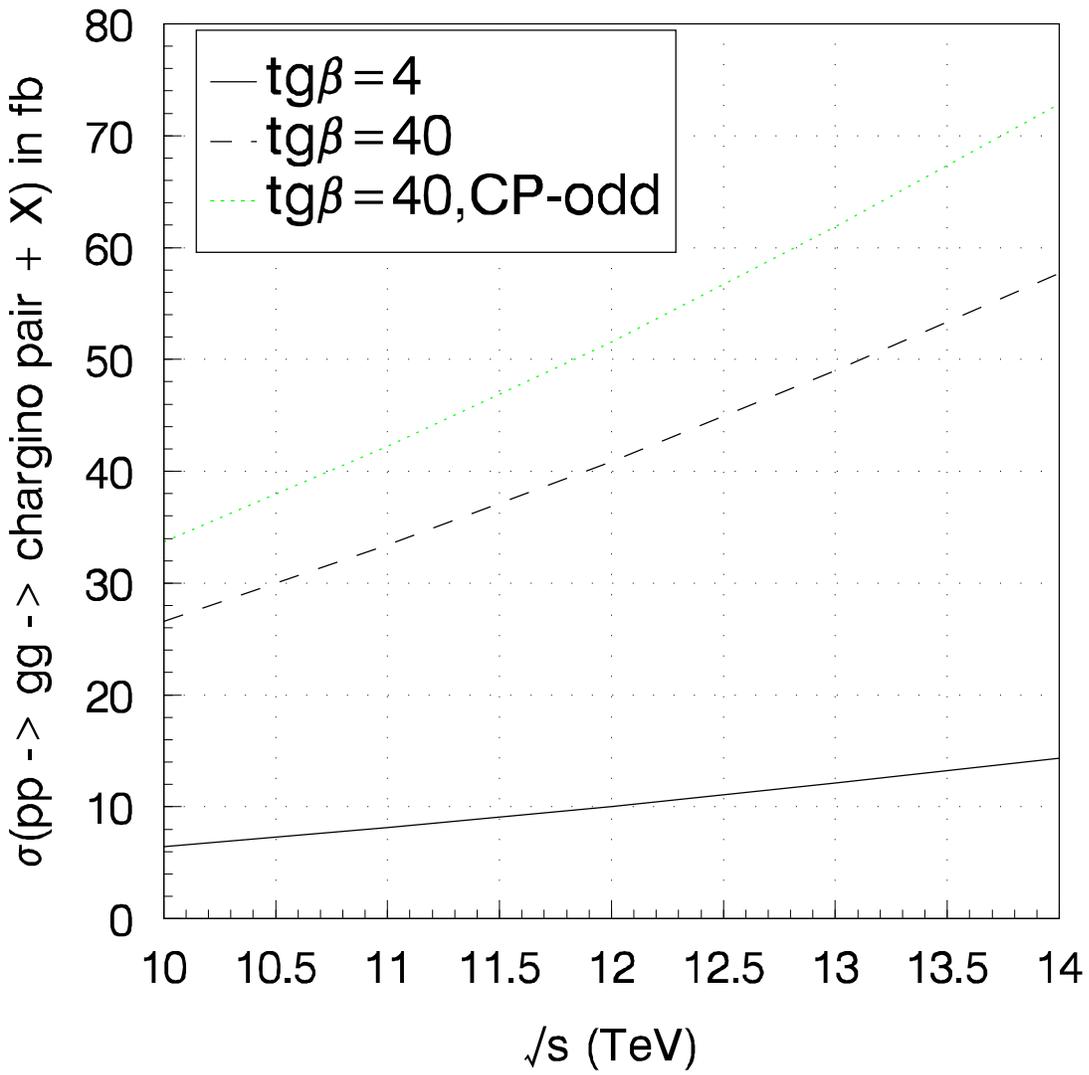
{\bf Fig.6} The cross section $\sigma$ of the higgsino-like chargino pair
     production process $pp \rightarrow gg \rightarrow \tilde{\chi}_1^{+}
     \tilde{\chi}_1^{-}+X$ versus the c.m.s energy of proton-proton collider
     $\sqrt{s}$. We take $m_{\tilde{\chi}^{+}_{1}}=165 GeV$,
     $m_{\tilde{\chi}^{+}_{2}}=750 GeV$, $m_{A}=250 GeV$, $\tilde{M}=200 GeV$
     (for the third generation) and the masses of the first and second
     generation squarks being 600 GeV. The full-line is for $\tan{\beta}=4$
     with all vanishing CP-odd phases, the dashed-line for $\tan{\beta}=40$
     with all vanishing CP-odd phases and the dotted-line for the case
     with $\tan{\beta}=40$, $\phi_{t}=\phi_{b}=\pi/4$ and other phase angels
     being zero.

\end{large}
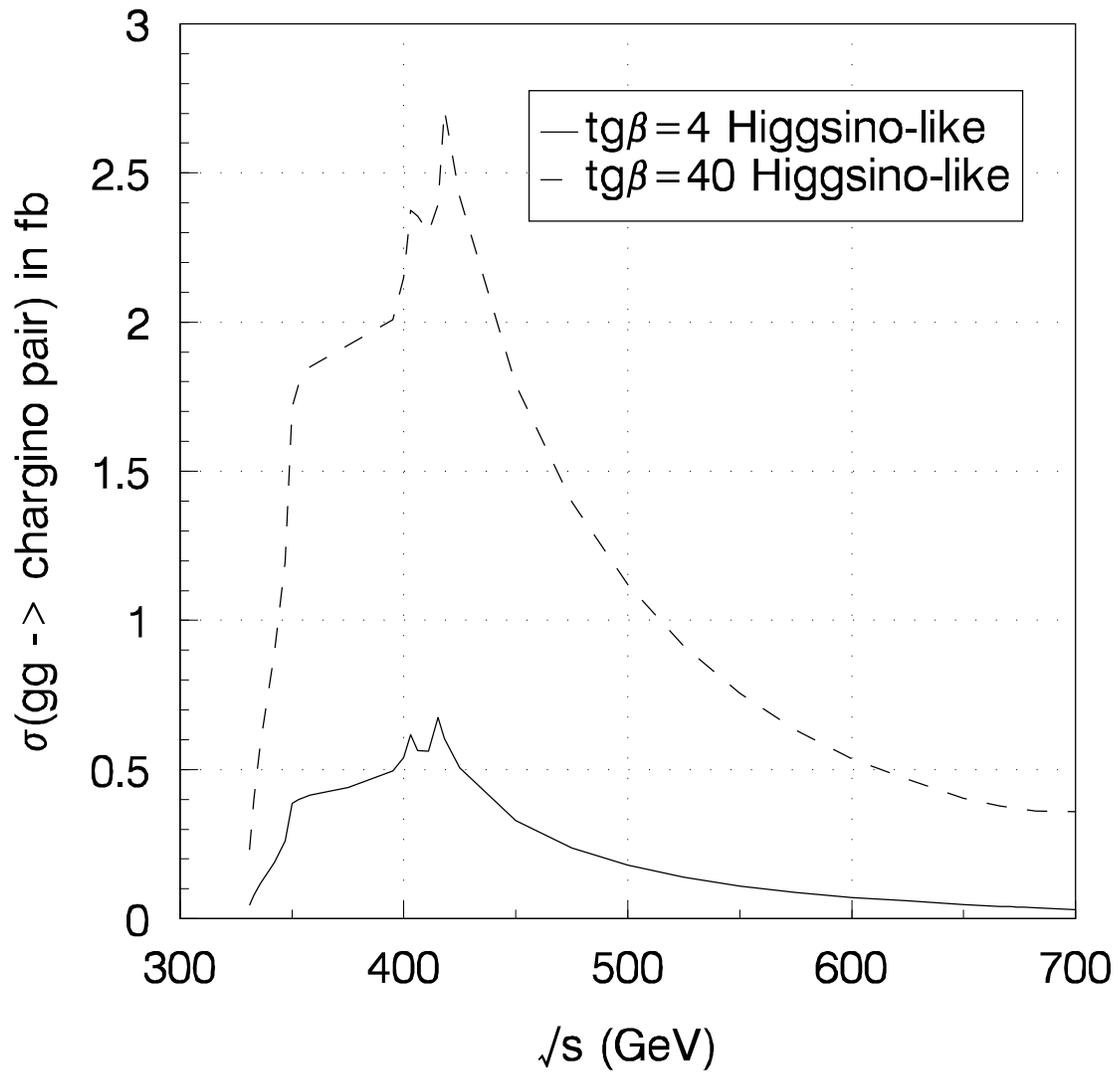
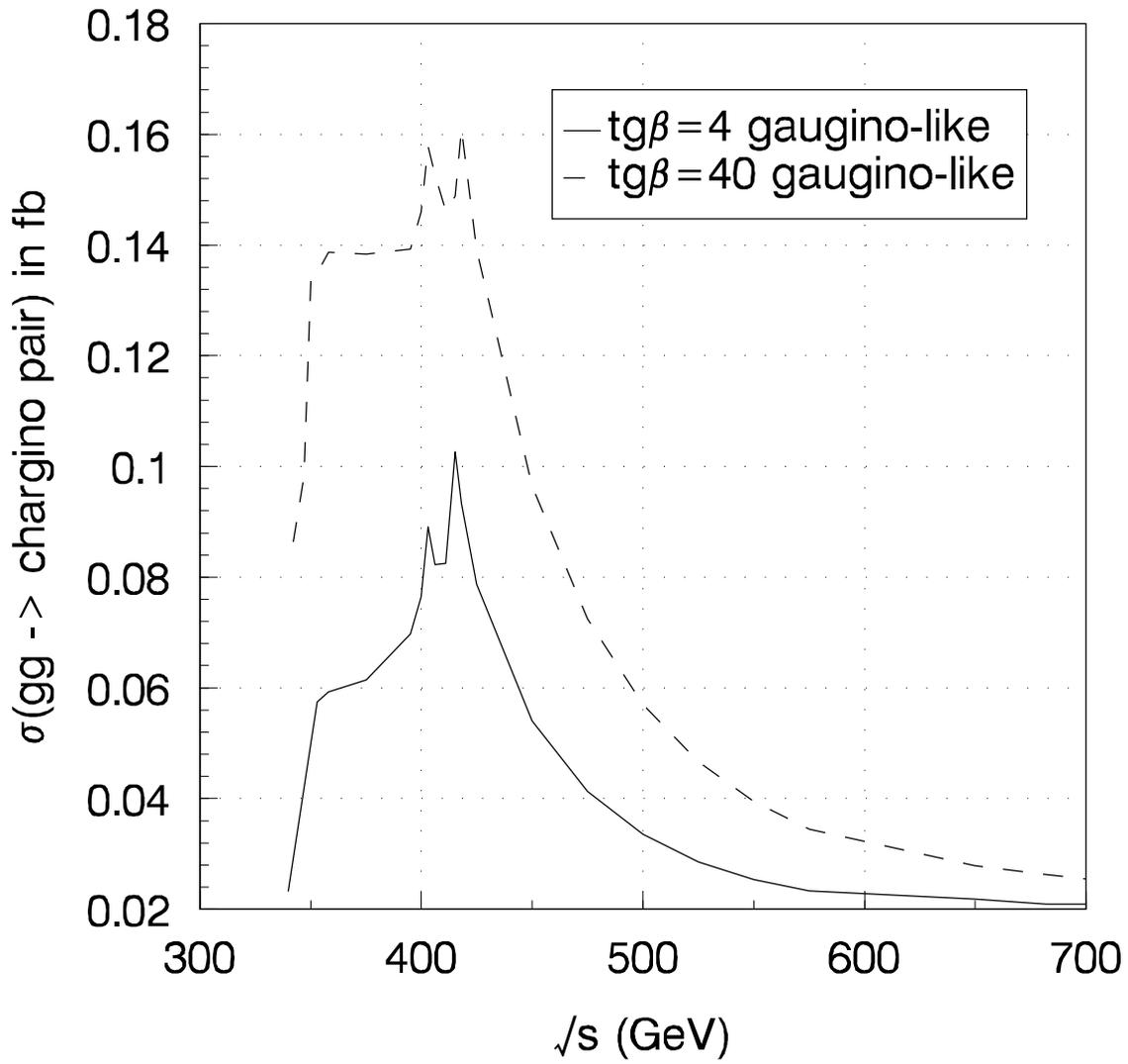
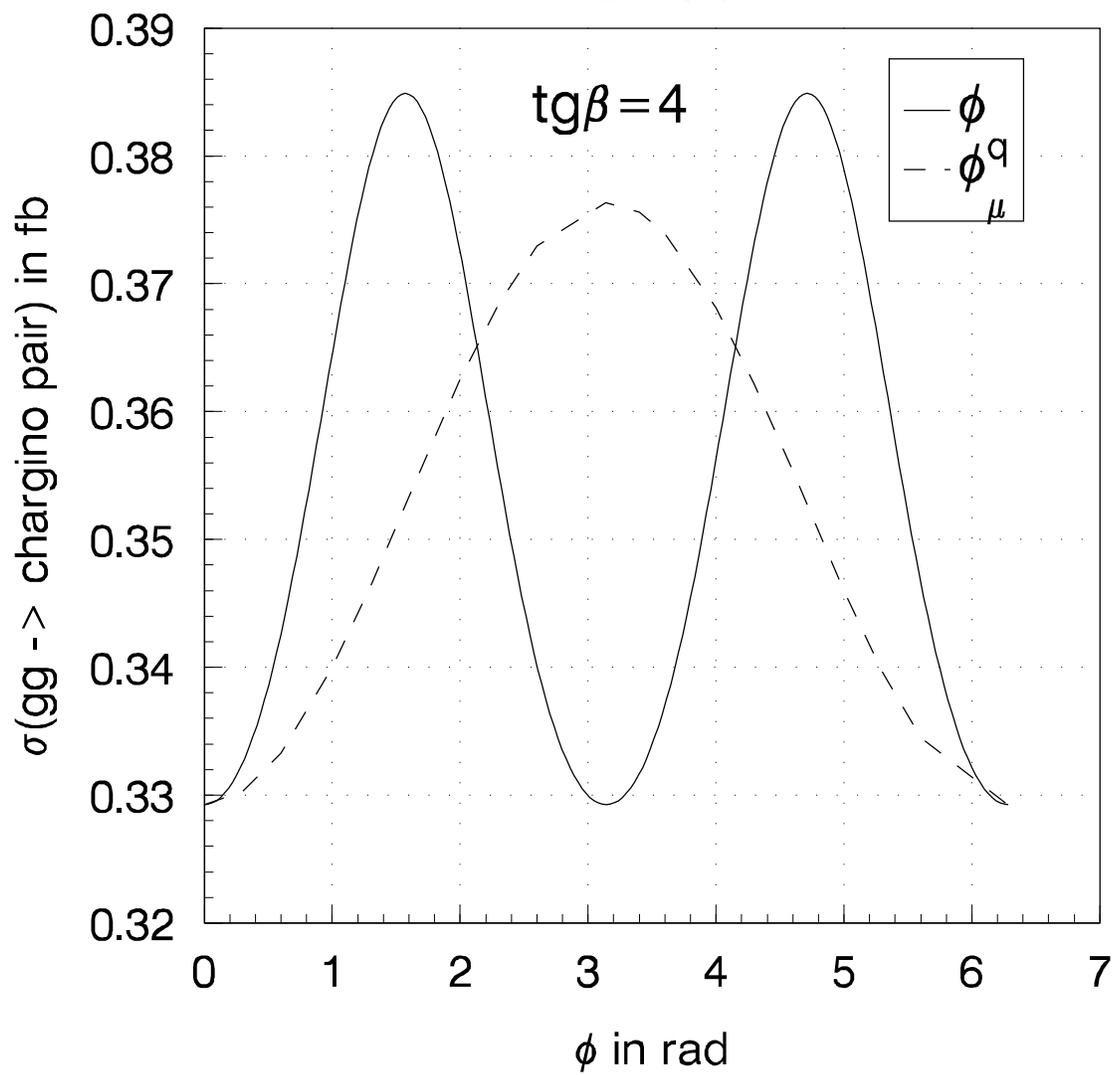
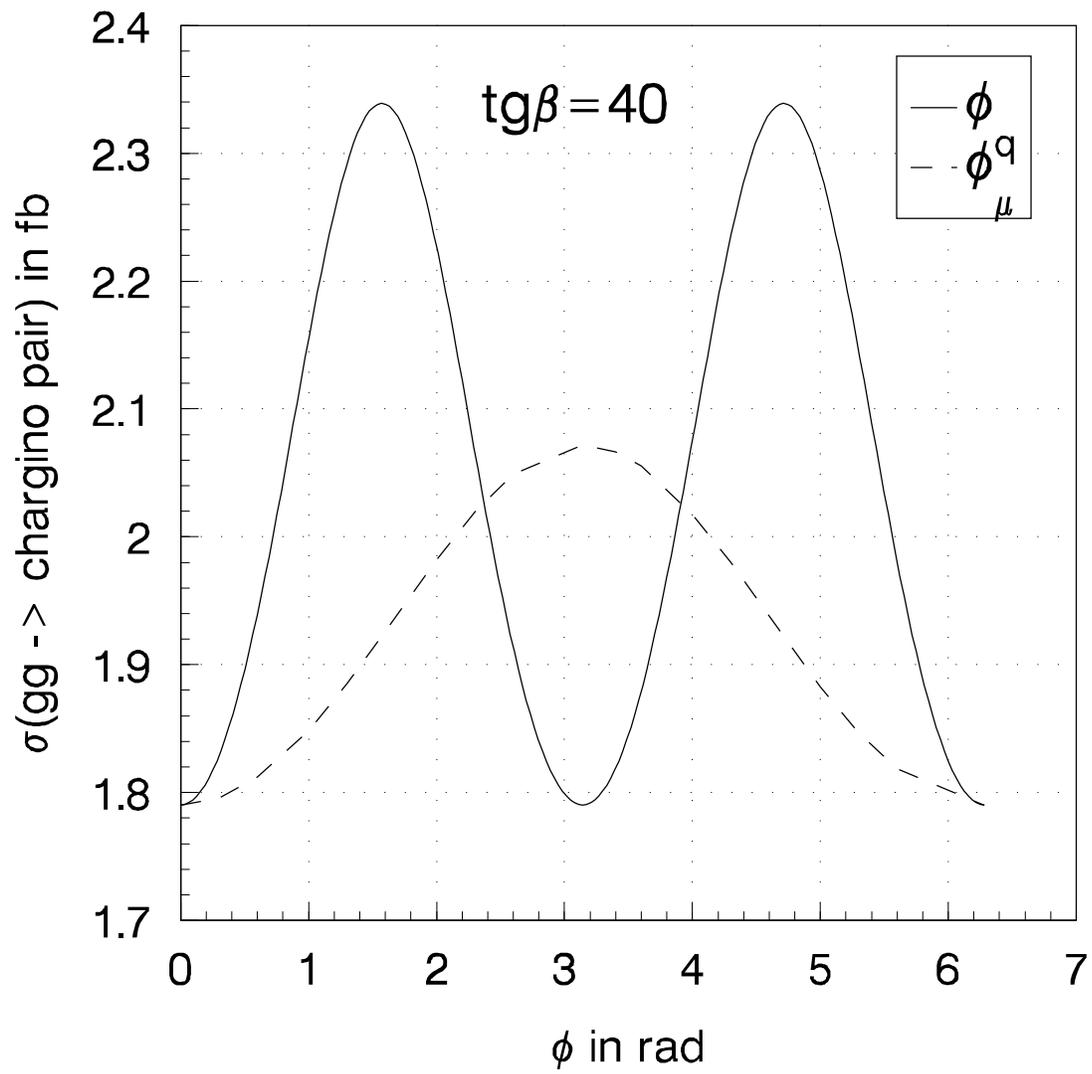
\end{document}